\documentclass[sigconf,screen]{acmart}

\usepackage{multirow}
\usepackage{graphicx}
\usepackage{diagbox}
\usepackage{subfigure}
\usepackage{xspace}
\usepackage{algorithm}
\usepackage{bbding}
\usepackage{threeparttable}
\usepackage{amsmath}
\usepackage{adjustbox}
\usepackage{listings}
\usepackage[T1]{fontenc}
\usepackage{semantic}
\usepackage{newfloat}
\usepackage{url}
\usepackage{amsmath}
\usepackage{framed}
\usepackage{verbatim}
\usepackage[utf8]{inputenc}
\usepackage[T1]{fontenc}
\usepackage{microtype}
\usepackage{cleveref}
\usepackage{wrapfig}
\usepackage{mathtools}
\usepackage{algorithm}
\usepackage{caption}
\usepackage{paralist}
\usepackage{enumitem}
\usepackage[noEnd=true,commentColor=blue]{algpseudocodex}
\usepackage{pifont}% http://ctan.org/pkg/pifont
\newcommand{\cmark}{\ding{51}}%
\newcommand{\xmark}{\ding{55}}%
\begin{document}

\newcommand{\bench}{SimdBench}
\newcommand{\tasknum}{136}
\newcommand{\llmnum}{18}
\newcommand{\anonymouslink}{https://anonymous.4open.science/r/SimdBench-1B3F/}

%\title{\bench{}: Benchmarking Intrinsic-Based Code-Vectorization for Large Language Models}
\title{\bench{}: Benchmarking Large Language Models for SIMD-Intrinsic Code Generation}

\author{Yibo He}
\orcid{0009-0002-1999-436X}
\affiliation{%
  \institution{Peking University}
  \department{Key Lab of HCST (PKU), MOE; SCS}
  \city{Beijing}
  \country{China}
}
\email{yibohe@pku.edu.cn}

\author{Shuoran Zhao}
\orcid{}
\affiliation{%
  \institution{Peking University}
  \department{School of EECS}
  \city{Beijing}
  \country{China}
}
\email{2300013126@stu.pku.edu.cn}

\author{Jiaming Huang}
\orcid{0000-0003-0798-1633}
\affiliation{%
  \institution{The Chinese University of Hong Kong, Shenzhen}
  \department{}
  \city{Shenzhen}
  \country{China}
}
\email{224040352@link.cuhk.edu.cn}

\author{Yingjie Fu}
\orcid{0000-0003-2574-9774}
\affiliation{%
  \institution{Peking University}
  \department{Key Lab of HCST (PKU), MOE; SCS}
  \city{Beijing}
  \country{China}
}
\email{yingjiefu@stu.pku.edu.cn}

\author{Hao Yu}
\orcid{0000-0002-3828-7612}
\affiliation{%
  \institution{The Hong Kong University of Science and Technology}
  \department{}
  \city{Hong Kong}
  \country{China}
}
\email{eehaoyusd@ust.hk}

\author{Cunjian Huang}
\orcid{0009-0008-4601-448X}
\affiliation{%
  \institution{DAMO Academy, Alibaba Group}
  \city{Hangzhou}
  \country{China}
}
\email{huangcunjian.huang@alibaba-inc.com}

\author{Tao Xie}
\authornote{Corresponding author.}
\orcid{0000-0002-6731-216X}
\affiliation{%
  \institution{Peking University}
  \department{Key Lab of HCST (PKU), MOE; SCS}
  \city{Beijing}
  \country{China}
}
\email{taoxie@pku.edu.cn}

\renewcommand{\shortauthors}{Yibo He et al.}

\begin{comment}
% Uncomment this for camera-ready version:
\author{
    \IEEEauthorblockN{
        Yibo He\textsuperscript{1},
        Shuoran Zhao\textsuperscript{2},
        Jiaming Huang\textsuperscript{3},
        Yingjie Fu\textsuperscript{1},
        Hao Yu\textsuperscript{4},
        Cunjian Huang\textsuperscript{5},
        Tao Xie\textsuperscript{1}
    }
    \IEEEauthorblockA{
        \textsuperscript{1}Key Lab of HCST (PKU), MOE; SCS, 
        \textsuperscript{2}Peking University, 
        \textsuperscript{3}CUHK-Shenzhen, 
        \textsuperscript{4}HKUST
        \textsuperscript{5}DAMO Academy, Alibaba Group
    }
    \IEEEauthorblockA{
        Emails: yibohe@pku.edu.cn, 2300013126@stu.pku.edu.cn, 224040352@link.cuhk.edu.cn,\\
        yingjiefu@stu.pku.edu.cn, eehaoyusd@ust.hk, huangcunjian.huang@alibaba-inc.com, taoxie@pku.edu.cn
    }
}
\end{comment}

\begin{abstract}
SIMD (Single Instruction Multiple Data) instructions and their compiler intrinsics are widely supported by modern processors to accelerate performance-critical tasks.
SIMD intrinsic programming, a trade-off between coding productivity and high performance, is widely used in the development of mainstream performance-critical libraries and daily computing tasks.
Large Language Models (LLMs), which have demonstrated strong and comprehensive capabilities in code generation, show promise in assisting programmers with the challenges of SIMD intrinsic programming.
However, existing code-generation benchmarks focus on only scalar code, and it is unclear how LLMs perform in generating vectorized code using SIMD intrinsics.
To fill this gap, we propose SimdBench, the first code benchmark specifically designed for SIMD-intrinsic code generation, comprising 136 carefully crafted tasks and targeting five representative SIMD intrinsics: SSE (x86 Streaming SIMD Extension), AVX (x86 Advanced Vector Extension), Neon (ARM Advanced SIMD Extension), SVE (ARM Scalable Vector Extension), and RVV (RISC-V Vector Extension).
We conduct a systematic evaluation (measuring both correctness and performance) of 18 representative LLMs on SimdBench, resulting in a series of novel and insightful findings.
Our evaluation results demonstrate that LLMs exhibit a universal decrease in pass@k during SIMD-intrinsic code generation compared to scalar-code generation.
Our in-depth analysis highlights promising directions for the further advancement of LLMs in the challenging domain of SIMD-intrinsic code generation.
SimdBench is fully open source at \url{https://anonymous.4open.science/r/SimdBench-1B3F/} to benefit the broader research community.

\end{abstract}




\maketitle

\section{Introduction}\label{sec:intro}

%\underline{S}ingle \underline{i}nstruction \underline{m}ultiple \underline{d}ata (SIMD) 
SIMD (Single Instruction Multiple Data) instructions are widely used in modern processors to accelerate performance-critical tasks by operating on multiple data items in parallel.
Given that coding assembly instructions is error-prone and extremely labor-intensive, programmers are required to vectorize their code implicitly (i.e., by compiler auto-vectorization~\cite{vectorization-nips19, vectorization-CGO06, vectorization-pldi06, vectorization-pldi16, vectorization-asplos21}) or explicitly (i.e., by SIMD intrinsics~\cite{Intel-Intrinsics-Guide, RVV-intrinsic, Neon, SVE}) to use SIMD instructions.
Despite substantial efforts toward effective auto-vectorization~\cite{vectorization-nips19, vectorization-CGO06, vectorization-pldi06, vectorization-pldi16, vectorization-asplos21}, compilers still struggle with vectorization failures and suboptimal optimizations due to fundamental limitations, such as insufficient compile-time information~\cite{limit_autov1, vectorization-eval-taco19}.

SIMD intrinsics~\cite{Intel-Intrinsics-Guide, RVV-intrinsic, Neon, SVE}, a type of built-in function provided by modern compilers, serve as a fundamental mechanism to achieve high efficiency in modern architectures.
%As an important way to optimize code with hardware-level features, SIMD intrinsics perform explicit vectorization by allowing programmers to manually specify SIMD instructions in high-level programming languages C/C++.
By enabling explicit vectorization, SIMD intrinsics allow programmers to directly specify SIMD instructions in high-level languages such as C and C++, thereby facilitating low-level code optimizations.
For fine-grained performance optimizations, SIMD intrinsics are widely used in the development of mainstream performance-critical libraries, such as OpenCV~\cite{OpenCV}, TensorFlow/XNNPACK~\cite{XNNPACK}, ONNX Runtime~\cite{ONNXRuntime}, and simdjson~\cite{simdjson}.
%Programmers can also use SIMD intrinsics to achieve high-performance solutions for daily tasks, e.g., LeetCode tasks~\cite{leetcodesolution}.
In addition, programmers can employ SIMD intrinsics to accelerate general-purpose tasks, such as solving algorithmic problems on platforms such as LeetCode~\cite{leetcodesolution}.

Compared to programming conventional scalar code, writing vectorized code using SIMD intrinsics presents a trade-off between coding productivity and high performance, falling into three main challenges.
(1) Complex interfaces and low readability. 
Low-level information, such as instruction-level operations, element types, and element width (or register width), is typically encoded into intrinsic names, leading to complexity and low readability.
(2) Manual data alignment and memory layout.
Programmers should manually load and store data from memory and align the data in each iteration of the loop to ensure correctness.
(3) Complex control flow and data flow.
Extra data operations and specific mechanisms, e.g., the masking mechanism~\cite{masking}, are used to implement vectorization of control flow and data flow, particularly in programs with irregular or data-dependent behaviors.

Large Language Models (LLMs), which have demonstrated strong and comprehensive capabilities in code generation~\cite{humaneval2021, science2022codegen, austin2021program}, are promising to assist programmers in solving the preceding challenges.
In recent years, proprietary LLMs such as GPT-3.5/GPT-4~\cite{openai2024gpt4}, Gemini~\cite{geminiteam2025geminifamily}, and open-source LLMs such as Qwen~\cite{qwen2025qwen25technicalreport} and DeepSeek-R1~\cite{deepseekai2025deepseekr1} have advanced at an unprecedented rate and achieved great success.
The code generation capabilities of these LLMs are evaluated by existing benchmarks (e.g., HumanEval~\cite{humaneval2021}) where LLMs are given a sequence of natural language tokens and are required to generate code snippets that pass the corresponding unit test cases.
However, none of the existing benchmarks~\cite{humaneval2021, apps2021, classeval2023, codereval2024, liu2023evalplus, austin2021program, ds1000, jimenez2024swebench, yang2024swebenchmultimodal, zhuo2025bigcodebench} have evaluated these LLMs on the challenging task of explicit vectorization via SIMD intrinsics, and it is unclear how LLMs perform in optimizing code with hardware-level features rather than generating only scalar code.

The limitations of current code generation benchmarks for evaluating LLMs on SIMD-intrinsic code generation can be categorized into three main aspects.
(1) The lack of domain-specific, vectorization-relevant tasks.
Considerable tasks in existing benchmarks lack requirements for batch processing or similar patterns and are therefore not suitable for vectorization.
(2) The absence of low-level implementation details required by SIMD intrinsics.
Low-level implementation details, such as element width in bytes, are absent from the prompts of existing benchmarks, which are designed either for highly abstract programming languages (e.g., Python)~\cite{classeval2023,humaneval2021,apps2021,codereval2024,jimenez2024swebench} that lack corresponding data type abstractions, or for multilingual tasks that rely on only default data types (e.g., \texttt{int} and \texttt{float}) in C/C++ tasks~\cite{multilingual2023,multilingual2024codegeex}.
(3) Insufficient test cases.
Most existing benchmarks focus on correctness testing while neglecting performance evaluation.
A few recent benchmarks~\cite{evalperf2024,EffiBench2024nips,Mercury2024nips,shypula2024pie, peng2025coffe} aim to address the neglect of performance evaluation, but only relevant metrics have been introduced, without improvements in testing on large-scale data or the establishment of frameworks to ensure performance precision and stability.

To address the preceding limitations, we propose \bench{}, the first benchmark specially designed for vectorization tasks for five representative types of SIMD intrinsics, i.e., intrinsics for SSE (x86 Streaming SIMD Extension)~\cite{Intel-Intrinsics-Guide}, AVX (x86 Advanced Vector Extension)~\cite{Intel-Intrinsics-Guide}, Neon (ARM Advanced SIMD Extension)~\cite{Neon}, SVE (ARM Scalable Vector Extension)~\cite{SVE}, and RVV (RISC-V Vector Extension)~\cite{RVV-intrinsic}.
\bench{} is distinguished by the following three features.
(1) Vectorization tasks for SIMD intrinsics. 
The \tasknum{} tasks in \bench{} are suitable for vectorization and are derived from two sources: (a) manually crafted based on six common operation types identified from typical SIMD intrinsic documents~\cite{Intel-Intrinsics-Guide, RVV-intrinsic, Neon, SVE}, and (b) filtered from one of the most widely used code generation benchmarks, HumanEval~\cite{humaneval2021}.
(2) Rich implementation details in function descriptions.
The details include element-wise operations, element types, element widths, and whether the inputs involve potentially unsafe behaviors (e.g., integer overflow).
(3) Comprehensive test cases for both correctness testing and performance testing.
Each task in \bench{} includes a test generator that produces valid, sufficient, and diverse test inputs for functional correctness testing. 
Additionally, performance test cases are incorporated for each task, designed to process large-scale data, and implemented using the Google Benchmark library~\cite{googlebenchmark} to ensure precise performance results.

We make the first experiment to evaluate \llmnum{} representative LLMs in the challenging generation scenario, i.e., generating explicitly vectorized code with SIMD intrinsics, and find three key results.
First, LLMs exhibit a universal decrease in the generation of SIMD-intrinsic code that is semantically correct compared to scalar-code generation.
Among the evaluated LLMs, DeepSeek-R1 performs the best, achieving an average pass@5 of 75.44\% for the five intrinsics.
Second, the valid code generated by LLMs using SIMD intrinsics results in further performance improvements compared to combining scalar code with compiler optimizations (including auto-vectorization) in a significant number of cases.
Third, the main obstacles to SIMD-intrinsic code generation currently are compilation errors related to the ``use of undeclared identifier'' (mainly for SVE and RVV) and logical bugs in the generated code (mainly for SSE, AVX, and Neon).
Our novel findings and in-depth analysis highlight promising directions for the further advancement of LLMs in the complex scenarios of generating SIMD-intrinsic code.

%\begin{comment}
In summary, this paper makes the following main contributions:
\begin{itemize}
%\begin{compactitem}
\item We introduce \bench{}, the first benchmark with \tasknum{} high-quality tasks specifically designed to generate vectorized code with SIMD intrinsics. 
\bench{} is available at \url{\anonymouslink}.
\item We conduct the first evaluation (including correctness and performance) on \llmnum{} representative closed-source or open-source LLMs on generating vectorized code implemented by five representative types of SIMD intrinsics.
\item We analyze the model capabilities based on our findings and highlight promising directions for advancing SIMD-intrinsic code generation with LLMs.
\end{itemize}
%\end{compactitem}
%\end{comment}
\section{Background}

\begin{figure}[t]
  \begin{center}
    \includegraphics[width=0.475\textwidth]{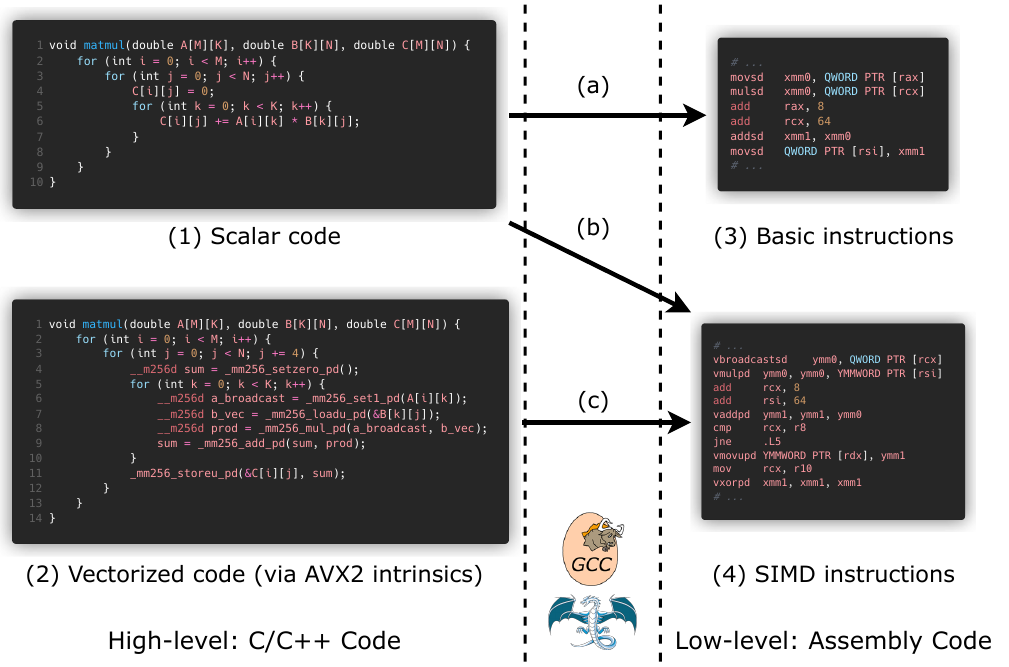}
  \end{center}
  %\vspace{-0.2cm}
  \caption{Examples of Matrix Multiplication Implementations. (a) Regular Compilation; (b) Implicit Vectorization via Compiler Optimizations; (c) Explicit Vectorization via SIMD Intrinsics.} 
  \label{fig:why-simd-intrinsic}
\Description[<short description>]{<long description>}
\end{figure}

% section 2.1
\subsection{SIMD Intrinsics}
As SIMD intrinsics are the main subject targeted by \bench{}, we discuss the essential domain knowledge of SIMD intrinsics in this section.
We first explain why modern architectures are designed with SIMD intrinsics in Section~\ref{sec2:whysimd}, and then introduce SIMD intrinsics in mainstream architectures in Section~\ref{sec2:mainsimd}.

\subsubsection{Why are modern architectures designed with SIMD intrinsics?}\label{sec2:whysimd}
With the advancement of information technologies, efficiently processing large-scale data has become increasingly important, leading to the growing significance and sophistication of SIMD (Single Instruction, Multiple Data) instructions in modern processors.
The key idea of SIMD is to compute multiple data elements simultaneously within a single instruction using multiple arithmetic logic units (e.g., Figure~\ref{fig:why-simd-intrinsic}(4)), rather than processing the elements sequentially with a single arithmetic logic unit (e.g., Figure~\ref{fig:why-simd-intrinsic}(3)).
Most modern architectures (e.g., x86, ARM, and RISC-V) support SIMD instructions.
As programming assembly code is error-prone and extremely labor-intensive, programmers typically adopt one of two main approaches to utilize SIMD instructions through high-level C/C++ code instead of writing low-level assembly.

(1) Implicit vectorization via compiler optimizations.
Modern compilers are typically implemented with auto-vectorization optimizations that generate SIMD instructions from scalar code (as shown in Figure~\ref{fig:why-simd-intrinsic}(b)), including loop-level vectorization~\cite{vectorization-pldi06,vectorization-pldi16} and superword-level parallelism (SLP)~\cite{vectorization-nips19,vectorization-asplos21}.
However, the effectiveness of compiler auto-vectorization depends on the compiler's capability to analyze programs for precise information at compile time~\cite{RefinedInputDegradedOutput}.
Existing research~\cite{limit_autov1, vectorization-eval-taco19} shows that the actual performance achieved by auto-vectorization is far from the architectural peak due to various obstacles, non-optimal optimizations, and inability to access certain information at compile time.

(2) Explicit vectorization via SIMD intrinsics.
SIMD intrinsics are built-in functions inside compilers, with functionality implemented by compilers.
SIMD intrinsics are designed to encapsulate SIMD instructions, allowing programmers to manipulate SIMD instructions like C/C++ functions (as shown in Figure~\ref{fig:why-simd-intrinsic}(c)).
Compilers relieve programmers of the tedious tasks involved in using SIMD instructions, such as register allocation and configuring control and status registers (CSRs).
Due to the preceding limitations of both manual assembly coding and compiler auto-vectorization, SIMD intrinsics are widely supported by modern architectures~\cite{Intel-Intrinsics-Guide, Neon, RVV-intrinsic} and are extensively used in the development of modern libraries~\cite{ONNXRuntime, OpenCV, simdjson, XNNPACK}.

\subsubsection{Mainstream SIMD intrinsics.}\label{sec2:mainsimd}

\begin{table*}[t]
    \centering
    %\small
    \footnotesize
    %\vspace{-0.2cm}
    \caption{SIMD-Instruction Extensions and Intrinsics in Mainstream Architectures.}
    \label{table:mainstream-simd}
\begin{tabular}{llllll}
\toprule
 & \textbf{SSE (x86)} & \textbf{AVX (x86)} & \textbf{Neon (ARM)} & \textbf{SVE (ARM)} & \textbf{RVV (RISC-V)} \\ \midrule
\textbf{Full Name} & Streaming SIMD Extensions & Advanced   Vector Extensions & Advanced   SIMD Extension & Scalable   Vector Extension & RISC-V Vector Extension \\ \hline
\textbf{Bit-width} & 128-bit & 256-bit, 512-bit (AVX-512) & 128-bit & Scalable: 128–2048 bits & Scalable \\ \hline
% & & & & & \\
%\multirow{-2}{*}{\textbf{Bit-width}} & \multirow{-2}{*}{128-bit} & \multirow{-2}{*}{\begin{tabular}[c]{@{}l@{}}256-bit, \\ 512-bit (AVX-512)\end{tabular}} & \multirow{-2}{*}{128-bit} & \multirow{-2}{*}{\begin{tabular}[c]{@{}l@{}}Scalable:\\ 128–2048 bits\end{tabular}} & \multirow{-2}{*}{Scalable} \\ \hline
\textbf{Length Agnostic} & \xmark & \xmark & \xmark & \cmark & \cmark \\ \hline
% & & & & & \\
%\multirow{-2}{*}{\textbf{Masked Ops}} & \multirow{-2}{*}{\xmark} & \multirow{-2}{*}{\begin{tabular}[c]{@{}l@{}}\cmark\\ (Mask in AVX-512)\end{tabular}} & \multirow{-2}{*}{\xmark} & \multirow{-2}{*}{\begin{tabular}[c]{@{}l@{}}\cmark\\ (Predicate Register)\end{tabular}} & \multirow{-2}{*}{\begin{tabular}[c]{@{}l@{}}\cmark\\      (Mask Register)\end{tabular}} \\ \hline
\textbf{Masked Ops} & \xmark & \cmark (Mask in AVX-512) & \xmark & \cmark (Predicate Register) & \cmark (Mask Register) \\ \hline
& & & & & \\
\multirow{-2}{*}{\textbf{Introduced in}}       & \multirow{-2}{*}{SSE (1999)} & \multirow{-2}{*}{\begin{tabular}[c]{@{}l@{}}AVX (2011),\\      AVX-512 (2013-2017)\end{tabular}}              & \multirow{-2}{*}{\begin{tabular}[c]{@{}l@{}}ARMv7/ARMv8\\      (2009)\end{tabular}} & \multirow{-2}{*}{\begin{tabular}[c]{@{}l@{}}ARMv8.2-A SVE\\      (2016)\end{tabular}}                       & \multirow{-2}{*}{\begin{tabular}[c]{@{}l@{}}RISC-V Vector   1.0\\      (2021)\end{tabular}} \\ \hline
\textbf{Instruction Ex.} & addps xmm, xmm & vaddps ymm, ymm, ymm & vadd.f32 q, q, q & faddp z, p/z, z & vadd.vv v, v, v \\ \hline
\textbf{Intrinsic Ex.} & \_mm\_add\_ps & \_mm256\_add\_ps & vaddq\_f32 & svaddp\_m & \_\_riscv\_vfadd\_vv\_f32m2 \\ \hline
  & & & & & \\
\multirow{-2}{*}{\textbf{Intrinsic Header}} & \multirow{-2}{*}{\begin{tabular}[c]{@{}l@{}}\textless{}xmmintrin.h\textgreater{},   \\      \textless{}emmintrin.h\textgreater{}\end{tabular}} & \multirow{-2}{*}{\textless{}immintrin.h\textgreater{}}                                                   & \multirow{-2}{*}{\textless{}arm\_neon.h\textgreater{}}                              & \multirow{-2}{*}{\textless{}arm\_sve.h\textgreater{}}                                                       & \multirow{-2}{*}{\textless{}riscv\_vector.h\textgreater{}} \\
\bottomrule
\end{tabular}
\end{table*}

We introduce five types of SIMD intrinsics in three mainstream modern architectures, x86, ARM, and RISC-V.
The five types of SIMD intrinsics are
(1) SSE series (e.g., SSE, SSE2, SSE3, SSE4.1, SSE4.2) for x86,
(2) AVX series (e.g., AVX, AVX2, AVX-512) for x86,
(3) Neon intrinsics for ARM,
(4) SVE intrinsics for ARM,
and (5) RVV intrinsics for RISC-V.
The overview of SIMD-instruction extensions and intrinsics in mainstream architectures is shown in Table~\ref{table:mainstream-simd}.

(1) x86 architecture.
Streaming SIMD Extensions (SSE) intrinsics and Advanced Vector Extensions (AVX) intrinsics are intrinsics for the x86 architecture. 
%that enable SIMD parallelism at the high-level programming languages.
SSE, which was introduced by Intel with the Pentium III processors in 1999, operates primarily on eight 128-bit registers known as \texttt{XMM0} through \texttt{XMM7}.
SSE is subsequently expanded by Intel to SSE2, SSE3, SSSE3, SSE4, etc.
AVX, a more recent extension introduced with Intel’s Sandy Bridge architecture, increases the register width to 256 bits (\texttt{\_\_m256}) and introduces a non-destructive three-operand format, which improves instruction scheduling and compiler optimization.
AVX-512 extends AVX with 512-bit support and introduces mask registers for selective execution and result blending.
Intrinsics for the x86 SIMD extensions are available in the Intel Intrinsics Guide~\cite{Intel-Intrinsics-Guide}.

(2) ARM architecture.
ARM architecture supports SIMD operations through two primary extensions: Neon~\cite{Neon} (i.e., Advanced SIMD Extension) and SVE~\cite{SVE} (i.e., Scalable Vector Extension).
Neon, introduced in the ARMv7-A architecture and extended in ARMv8-A, provides fixed-width 128-bit SIMD operations. 
Neon supports integer and floating-point arithmetic, vectorized memory operations, and type conversions through a set of C/C++ intrinsics.
For example, the intrinsic \texttt{vaddq\_f32} performs an element-wise addition of two 128-bit vectors containing four single-precision floating points.
SVE, introduced with ARMv8.2-A, extends SIMD capabilities beyond fixed vector widths by supporting scalable vectors. 
Unlike Neon, SVE does not define a fixed vector size; instead, the size can vary from 128 bits to 2048 bits in 128-bit increments depending on the hardware implementation.
SVE intrinsics allow predicated operations (using predicate registers), vector-length agnostic programming, and advanced features such as gather/scatter and complex number arithmetic.
Intrinsics for the ARM SIMD extensions are available in the Neon document~\cite{Neon} and the SVE document~\cite{SVE}.

(3) RISC-V architecture.
RISC-V supports SIMD operations mainly through the RISC-V Vector Extension (RVV).
In contrast to fixed-width SIMD architectures such as SSE and AVX, RVV adopts a vector-length agnostic (VLA) programming model, where the vector length (i.e., \texttt{VLEN}) is determined by the hardware and typically ranges from 128 to 4096 bits.
RVV intrinsics follow a unified naming convention prefixed with \texttt{\_\_riscv\_}, which encodes both semantic meaning and implementation details.
For example, the intrinsic \texttt{\_\_riscv\_vfadd\_vv\_f32m2} performs element-wise addition between two vectors of single-precision floating-point numbers.
Additionally, RVV intrinsics support masking operations to facilitate vectorized control flow.
A comprehensive list of RVV intrinsics is available in the official RVV-intrinsic documentation~\cite{RVV-intrinsic}.

% section 2.2
\subsection{Code Generation Benchmarks}

Benchmarks for code generation play a crucial role in evaluating and comparing the effectiveness of various LLMs and coding techniques.
Existing benchmarks for code generation can be divided into two main categories, general-purpose code benchmarks and domain-specific code benchmarks.
General-purpose code benchmarks, such as HumanEval~\cite{humaneval2021}, MBPP~\cite{austin2021program}, and APPS~\cite{apps2021}, typically focus on generating general-purpose code snippets or solving programming challenges across various languages~\cite{multilingual2023,multilingual2024codegeex}. 
Domain-specific code benchmarks emphasize specific domains, such as data processing~\cite{ds1000}, machine learning~\cite{li2025tritonbenchbenchmarkinglargelanguage}, and unit testing~\cite{jain2025testgenevalrealworldunit}. 

However, these benchmarks lack coverage of low-level features, such as SIMD optimizations, SPMD (Single Program Multiple Data, e.g., Intel ISPC~\cite{Intel-SPMD}) optimizations, and cross-architecture portability, which are critical for evaluating code generation in performance-sensitive and system-level programming contexts. 
Consequently, there is a growing need for more specialized benchmarks that capture these advanced code generation scenarios.
To address this gap, we propose \bench{}, a benchmark designed to target one of the most fundamental features: SIMD-intrinsic code generation.
\section{Benchmark Construction}\label{sec:approach}

In this section, we introduce \bench{}, the first benchmark that is specifically designed to evaluate the capabilities of LLMs in generating vectorized code with SIMD intrinsics.
\bench{} consists of \tasknum{} high-quality tasks, each targeting five representative SIMD intrinsics: SSE (x86 Streaming SIMD Extension)~\cite{Intel-Intrinsics-Guide}, AVX (x86 Advanced Vector Extension)~\cite{Intel-Intrinsics-Guide}, Neon (ARM Advanced SIMD Extension)~\cite{Neon}, SVE (ARM Scalable Vector Extension)~\cite{SVE}, and RVV (RISC-V Vector Extension)~\cite{RVV-intrinsic}.
We present the overview of \bench{} with an example task in Section~\ref{sec:task-overview}, the construction of tasks in Section~\ref{sec:task-construction}, and the construction of test cases in Section~\ref{sec:test-construction}.

\subsection{Overview of \bench{}}\label{sec:task-overview}

% \bench{}: Benchmarking Large Language Models for SIMD-Intrinsic Code Generation

\begin{figure}[t]
  \begin{center}
    \includegraphics[width=0.4\textwidth]{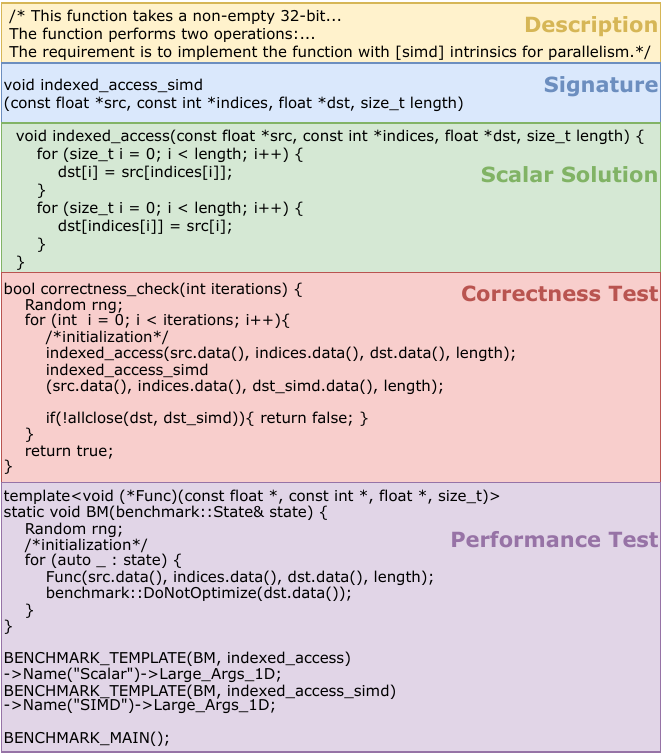}
  \end{center}
  \caption{Overview of a Task in \bench{}.} 
  \label{fig:task-overview}
\Description[<short description>]{<long description>}
\end{figure}

Figure~\ref{fig:task-overview} shows an example of a task in \bench{}.
Each coding task in \bench{} consists of a functional description in natural language, a signature for the target function, a correctness test case for validating the generated code, and a performance test case for evaluating the efficiency of the generated code.
Other miscellaneous information is manually annotated and provided, such as the entry points for the scalar function (e.g., \texttt{indexed\_access}) and the vectorized function (e.g., \texttt{indexed\_access\_simd}).
The example in Figure~\ref{fig:task-overview} aims to retrieve data from non-contiguous memory locations (i.e., gather operation) and write data to non-contiguous memory locations (i.e., scatter operation) based on the same array of indices.
This task is well-suited for acceleration through vectorization using SIMD intrinsics.

In the typical workflow, LLMs generate vectorized code snippets based on input prompts containing processed functional descriptions, where the token ``[simd]'' is replaced with the corresponding full name of SIMD intrinsics.
The generated code snippet is verified using a correctness test case, which conducts differential testing by comparing the results of the generated code with the results of the canonical scalar solution.
If a generated code snippet passes the correctness check, the snippet is evaluated using a performance test case, where speedup is calculated by comparing the performance of the generated code against the canonical scalar solution as the baseline, using the same machine, compiler, and compiler options.

In our experiments, we focus on the code generation task using only processed functional descriptions as input prompts to the LLMs. 
Other tasks, such as translating scalar code to vectorized code, and advanced prompting techniques, such as Chain-of-Thought (CoT) and Retrieval-Augmented Generation (RAG), are beyond the scope of this paper but represent promising directions for future research in SIMD-intrinsic code generation.

\subsection{Task Construction}\label{sec:task-construction}

%The tasks in \bench{} are from two sources: (1) hand-written tasks based on basic operations refined from intrinsic documents, and (2) revised tasks based on HumanEval~\cite{humaneval2021} and HumanEval-CPP~\cite{multilingual2024codegeex} (i.e., C++ version of HumanEval).
The tasks in \bench{} are derived from two sources: (1) hand-crafted tasks based on basic operations refined from the intrinsic documentation, and (2) modified tasks based on HumanEval~\cite{humaneval2021} and its C++ variant, HumanEval-CPP~\cite{multilingual2024codegeex}.

\textbf{Hand-crafted tasks based on SIMD-intrinsic documentation.}
Hand-crafted tasks based on basic operations are designed to cover common operations and data types that are supported by all five target SIMD intrinsics.
To achieve this goal, we refer to the documentation for each of the five SIMD intrinsics and refine six basic operations that are commonly supported across SIMD intrinsics for various architectures.
The six basic operations are as follows: 
(1) data movement operations, which transfer data between a source memory address and a destination address; 
(2) integer operations, which perform arithmetic operations on data of integer types; 
(3) floating-point operations, which perform arithmetic operations on data of floating-point types; 
(4) comparison operations, e.g., $=, \neq, >, <, $ and their combinations with arithmetic operations; 
(5) logical operations, including AND, OR, XOR, NOT operations, along with their combinations with other operations; 
and (6) reinterpret cast operations, which reinterpret data types.
We construct around 10 tasks for each basic operation, resulting in a total of 62 hand-crafted tasks in \bench{}.

The data types used in \bench{} are supported by all five evaluated SIMD intrinsics, including signed and unsigned integer types (8-, 16-, 32-, and 64-bit), floating-point types (32- and 64-bit), as well as the \texttt{bool} and \texttt{string} types.
We exclude types that are not supported by all architectures, such as integers in 128-bit and low-precision floating-point types.
We also exclude data types with indeterminate sizes, such as \texttt{short} and \texttt{long}, whose sizes differ between 32-bit and 64-bit architectures.

The hand-crafted tasks are constructed in the following four steps.
(1) We implement a code snippet with vectorizable loops based on a basic operation and one or more data types that are infrequent in the constructed tasks.
(2) We write a functional description to describe the functionality of the code snippet. 
The function description includes the input parameters and their types, identification of potential undefined behaviors (e.g., accessing a NULL pointer), operations performed, output, and a prompt for utilizing SIMD intrinsics for enabling parallelism.
(3) We create a correctness test case and a performance test case for this task. The details of test construction are discussed in Section~\ref{sec:test-construction}.
(4) We analyze the frequency of basic operations and data types covered in the constructed tasks to guide the development of new tasks.

\textbf{Modified tasks based on HumanEval and HumanEval-CPP.}
Since the scalar-code generation capabilities of LLMs have been rigorously evaluated using the well-established HumanEval benchmark~\cite{humaneval2021} and its variants~\cite{liu2023evalplus,multilingual2024codegeex,evalperf2024}, we incorporate tasks from HumanEval and its C++ variant (i.e., HumanEval-CPP) to enhance the diversity of tasks in \bench{}.

As discussed in Section~\ref{sec:intro}, considerable tasks in existing code generation benchmarks lack requirements for batch processing and are unsuitable for SIMD-intrinsic code generation.
For this reason, we excluded tasks from HumanEval that are unsuitable for SIMD-intrinsic code generation based on the following principles:
(1) sequential tasks, i.e., tasks without loops or with loops having very limited iteration counts;
(2) tasks with inherently sequential loops or recursive functions (e.g., Fibonacci);
(3) tasks with loops involving data structures with non-contiguous and unpredictable memory layouts (e.g., a string array with an uncertain number of \texttt{push} and \texttt{pop} operations);
(4) tasks with loops involving control flow with function or subroutine calls (except SIMD intrinsics);
and (5) tasks with loops involving third-party libraries or function calls that are not supported by C/C++.

After the preceding filtering step, we select 74 tasks from the 164 tasks in HumanEval that are suitable for SIMD-intrinsic code generation. 
Among the tasks in HumanEval, 41 belong to type (1), 9 to type (2), 31 to type (3), 8 to type (4), and 1 to type (5).
Detailed task IDs for the filtered tasks are provided in our open-source code repository~\cite{simdbench}.

We modify the selected problems to match the format of hand-crafted tasks, ensuring that all tasks in \bench{} are compatible with the same evaluation framework.
(1) We revise the imprecise problem descriptions and augment the functional descriptions in HumanEval-CPP with prompts that enable parallelism using SIMD intrinsics.
As prior work~\cite{liu2023evalplus} demonstrates, HumanEval contains imprecise problem descriptions; we find that HumanEval-CPP exhibits similar issues, particularly in C++-specific implementation details.
(2) We construct a correctness test case and a performance test case for each task.
The original test cases in HumanEval-CPP are reused as special values during the first few iterations when initializing test inputs in the correctness test cases.
(3) We moderately expand the domain of problem input (e.g., the length of an input array) in functional descriptions to enable performance evaluation using larger-scale data.

\textbf{Prompt processing before generating samples.}
The functional descriptions in \bench{} are prompt templates and should be preprocessed before being used to generate SIMD-intrinsic code for specific architectures.
The token ``[simd]'' is replaced with the corresponding full name of the SIMD intrinsics when preprocessing prompts.
We also generate a scalar-code implementation as a baseline during correctness evaluation.
To reuse the functional descriptions and test cases, we remove the parallelism-enabling prompts and replace the entry point of the vectorized function with that of the scalar function in the functional description when generating scalar code.
Before evaluating the generated scalar code, we restore the entry point to that of the vectorized function, resulting in a scalar implementation that uses the vectorized function's name.

\begin{table}[t]
  \centering
  \scriptsize
  \caption{Summary Statistics of HumanEval-CPP~\cite{multilingual2024codegeex} and \bench{} (Function-level C/C++ Programming). \#Char.: Characters. Number of Covered Main Types by Parameters and Return Types (Second Table).}
  \label{tab:statistics-bench}
  \begin{minipage}{0.475\textwidth}
    \centering
\begin{tabular}{llcc|cc|cc}
\toprule
%\multirow{2}{*}{Benchmark} 
 & \multirow{2}{*}{\#Task} & \multicolumn{2}{c|}{Test (Avg.)} & \multicolumn{2}{c|}{Prompt (Avg.)} & \multicolumn{2}{c}{Solution (Avg.)} \\ \cline{3-4} \cline{5-6} \cline{7-8}
 &     & \#N & \#Char. & \#Char. & \#Line & \#Char. & \#Line \\ \midrule
HumanEval-CPP  & 164 & 7.7 & 503.3 & 481.2 & 16.3 & 259.2  & 10.6  \\ \hline
SimdBench  & \tasknum{} & $+\infty$ & 1813.8 & 668.4 & 17.9 & 318.7 & 10.7  \\ \bottomrule
\end{tabular}
  \end{minipage}%

  \begin{minipage}{0.475\textwidth}
    \centering
    %\caption{Summary Statistics for Main Data Types.}
    %\label{tab:statistics-2}
    % \vspace{-0.2cm}
\begin{tabular}{lcccc|cccc|cc|c}
\toprule
Type & \multicolumn{4}{c|}{\#int} & \multicolumn{4}{c|}{\#uint} & \multicolumn{2}{c|}{\#float} & \#bool \\ \cline{2-12}
Bits & 64    & 32   & 16   & 8   & 64    & 32    & 16   & 8   & 64   & 32   & 1  \\ \midrule
HumanEval-CPP & 2 & \textbf{107} & 0 & 0 & 0 & 0 & 0 & 0 & 2 & 26 & \textbf{27} \\
SimdBench & \textbf{21} & 62 & \textbf{9} & \textbf{9} & \textbf{22} & \textbf{12} & \textbf{9} & \textbf{8} & \textbf{38} & \textbf{43} & \textbf{27} \\ \bottomrule
\end{tabular}
  \end{minipage}
\end{table}

\subsection{Test Construction}\label{sec:test-construction}
For each task in \bench{}, a correctness test case and a performance test case are provided for a comprehensive evaluation.

\textbf{Correctness test.} 
Correctness test cases use differential testing to validate the code generated by LLMs by comparing the results between the generated code and the canonical scalar solution.
Figure~\ref{fig:task-overview} presents an example of a correctness test case.
%Correctness test cases first initialize the inputs of tested functions, get results of canonical scalar implementations and SIMD-intrinsic implementations separately, and compare the results to validate correctness.
Each test case initializes the inputs, executes both the scalar and SIMD-intrinsic implementations, and compares their outputs to check for correctness.
Input initialization falls into three categories: 
(1) special values, including preset corner cases and reused inputs; 
(2) randomized values within valid ranges, such as array lengths, to ensure correctness and safety;
and (3) fully randomized values, such as the individual elements of safe arrays.
The correctness test case repeats the preceding iteration $N$ (the parameter of correctness test cases) times and returns \texttt{true} if and only if all iterations pass.
During the first few iterations, test inputs are initialized with special values; in the remaining iterations, test inputs are initialized randomly.

We perform correctness validation using correctness test cases in the following three steps.
(1) To detect whether the generated code lacks intrinsics (i.e., the LLM ignores the prompts and generates scalar code instead of SIMD-intrinsic code), we compile the generated code together with the correctness test case \textbf{without} the required headers and compiler options.
If the code compiles successfully under these conditions, we label the sample as ``failed due to no intrinsic''.
(2) We compile the generated code and the corresponding test case with the required headers and compiler options.
(3) We execute the resulting binary and check whether the result is \texttt{true}.
A sample is labeled as passed if and only if the generated code in the sample contains SIMD intrinsics, compiles and executes successfully, and returns \texttt{true}.

\textbf{Performance test.}
Performance test cases are implemented using the Google Benchmark library~\cite{googlebenchmark}, which is specifically developed to measure and evaluate the performance of small C/C++ code segments.
Figure~\ref{fig:task-overview} presents an example of a performance test case.
In each performance test case, a function template is developed to initialize the inputs of the tested function and invoke the function.
Then the function template is invoked twice: once for the generated SIMD-intrinsic code and once for the canonical scalar solution.
In each execution of the function template, the code snippet under test is executed repeatedly multiple times to record the stable execution time with high precision (typically in nanoseconds).

The initialization of the test input in performance test cases is similar to, but distinct from, that in correctness test cases.
We analyze the time complexity of the solution code and categorize the test input into two types: (1) stable input (e.g., \texttt{length} in Figure~\ref{fig:task-overview}), which affects the time complexity and should remain constant during iterations in the execution of a function template, and (2) arbitrary input (e.g., values in \texttt{src} and \texttt{dst} in Figure~\ref{fig:task-overview}), which does not affect the time complexity and can be various.
We initialize stable inputs with predefined seeds and arbitrary inputs with random seeds.
Additionally, the range of array lengths is predefined and larger than that in the correctness test cases to conduct performance evaluation with large-scale data.
This design ensures that the performance difference is attributed to the code under test, rather than the initialization of the test inputs.

The implementation of performance test cases ensures the stability, precision, and reliability of performance results.

\textbf{Statistics.}
Table\ref{tab:statistics-bench} presents an overview comparison between HumanEval-CPP and \bench{}.
As shown in Table\ref{tab:statistics-bench}, HumanEval-CPP covers only limited data types, but \bench{} covers all main data types for calculation.
\bench{} includes more complex test cases than HumanEval-CPP, using richer input scenarios rather than relying solely on multiple lines of assertions.

\subsection{Evaluation Metrics}\label{sec:metric}
Three metrics are used in \bench{}: pass@k~\cite{humaneval2021}, speedup, and efficient@k~\cite{peng2025coffe}.
Pass@k represents the mean probability, computed across all problems in the benchmark, that one or more solutions out of $k$ samples successfully pass the correctness test case.

%\begin{equation} \label{eq:passatk}
\[
\mathrm{pass}@k = \mathbb{E}_{\mathrm{task}} \left[ 
  1 - \frac{\binom{n - c}{k}}{\binom{n}{k}} 
\right]
\]
%\end{equation}

We use the speedup and efficient@k metrics for performance evaluation.
Speedup measures the degree of acceleration achieved by each valid sample.
For each valid sample, the performance test case is executed with multiple preset input arguments, and each argument is run multiple times, yielding multiple speedup measurements.
We sort the initial speedup results, exclude the top 20\% and bottom 20\% to avoid bias, and calculate the average of the remaining results ($s_i$) to obtain the final speedup result of a valid sample (as shown in the following equation).
As averaging speedup results across multiple problems is confusing~\cite{evalperf2024}, we use efficient@k, which is derived from the previous work~\cite{peng2025coffe}, to represent the mean probability computed across all problems in the benchmark, that one or more solutions out of $k$ samples are valid and achieve a positive performance impact (i.e., $speedup > 1.0$).

%\begin{equation} \label{eq:speedup}
\[
%speedup = \frac{T_{scalar}}{T_{simd}}，
speedup = \frac{\sum\limits_{i=k+1}^{n-k}{s_{i}}}{n-2k}, k = round(0.2*n)
%speedup = (\sum\limits_{i=k}^{n-k}{s_{i}}) / (n-2k), k = round(0.2*n)
\]
%\end{equation}

%\begin{equation} \label{eq:efficientatk}
\[
\mathrm{efficient}@k = \mathbb{E}_{\mathrm{task}} \left[ 
  1 - \frac{\binom{n - c_{f}}{k}}{\binom{n}{k}}
\right], c_{f} = 
\begin{cases}
n(speedup > 1.0) \\
n(speedup \leq 1.0) \\
\end{cases}
\]
%\end{equation}

\begin{table*}[t]
\centering
\caption{Evaluated LLMs in Our Experiments.} \label{table:models}
\begin{tabular}{lllc|lllc}
\toprule
\textbf{Model} & \textbf{Tag} & \textbf{Org.} & \textbf{Open-Source}  & \textbf{Model} & \textbf{Tag} & \textbf{Org.} & \textbf{Open-Source} \\ \midrule
Claude-3-5-Haiku~\cite{claude}  & 20241022 & Anthropic & \xmark  & GPT-4o-mini~\cite{openai-platform} & 2024-07-18 & OpenAI & \xmark  \\
Claude-3-5-Sonnet~\cite{claude} & 20241022 & Anthropic & \xmark  & Grok-3~\cite{grok} & - & xAI & \xmark  \\
Codestral~\cite{Codestral} & 25.01 & Mistral & \xmark  & Grok-3-Beta~\cite{grok} & - & xAI & \xmark  \\
DeepSeek-V3~\cite{deepseekai2024deepseekv3technicalreport}  & 0324 & DeepSeek & \cmark (671B) & Mistral-Large~\cite{Mistral-Large} & 24.11 & Mistral & \cmark (123B) \\
DeepSeek-R1~\cite{deepseekai2025deepseekr1} & 0528 & DeepSeek & \cmark (671B) & Qwen-Coder-Plus~\cite{hui2024qwen25codertechnicalreport}  & 2024-11-06 & Alibaba & \xmark  \\
Gemini-2.0-Flash~\cite{gemini-api}  & - & Google & \xmark  & Qwen-Coder-Turbo~\cite{hui2024qwen25codertechnicalreport} & 2024-09-19 & Alibaba & \xmark \\
Gemini-2.5-Flash~\cite{gemini-api}  & 05-20 & Google & \xmark  & Qwen2.5-Max~\cite{qwen2025qwen25technicalreport} & 2025-01-25 & Alibaba & \xmark  \\
GPT-3.5-Turbo~\cite{openai-platform} & 0125 & OpenAI & \xmark  & Qwen2.5-Plus~\cite{qwen2025qwen25technicalreport} & 2025-01-25 & Alibaba & \xmark  \\
GPT-4o~\cite{openai-platform} & 2024-11-20 & OpenAI & \xmark  & Qwen2.5-Turbo~\cite{yang2025qwen251mtechnicalreport} & 2025-02-11 & Alibaba & \xmark \\ \bottomrule
\end{tabular}
\end{table*}

\begin{table*}[t]
\centering
%\footnotesize
\small
%\vspace{-0.2cm}
\caption{Correctness-Evaluation Results of LLMs on \bench{} (\%).}
\label{table:pass@k_results}
%\begin{tabular}{lllllllllllll}
\begin{tabular}{l|cc|cc|cc|cc|cc|cc}
\toprule
\textbf{}        & \multicolumn{2}{c|}{\textbf{Scalar}} & \multicolumn{2}{c|}{\textbf{SSE}} & \multicolumn{2}{c|}{\textbf{AVX}} & \multicolumn{2}{c|}{\textbf{Neon}} & \multicolumn{2}{c|}{\textbf{SVE}} & \multicolumn{2}{c}{\textbf{RVV}} \\
\midrule
Model   & Pass@1 & Pass@5 & Pass@1 & Pass@5 & Pass@1 & Pass@5 & Pass@1 & Pass@5 & Pass@1 & Pass@5 & Pass@1 & Pass@5 \\
\midrule
% Model          &    scalar     &      SSE      &      AVX      &     Neon      &      SVE      &      RVV      \\
Claude-3-5-Haiku & 78.97 & 80.88 & 20.44 & 32.35 & 24.12 & 36.76 & 23.82 & 35.29 & 13.09 & 23.53 & 13.97 & 21.32 \\
Claude-3-5-Sonnet& 81.32 & 86.03 & 31.18 & 44.85 & 38.09 & 52.21 & 33.38 & 47.06 & 12.65 & 23.53 & 01.76 & 05.15 \\
Codestral        & 80.44 & 83.09 & 06.76 & 11.03 & 06.62 & 13.24 & 04.12 & 11.03 & 06.18 & 10.29 & 03.68 & 09.56 \\
DeepSeek-V3      & 85.88 & 87.50 & 28.24 & 40.44 & 36.47 & 51.47 & 31.32 & 45.59 & 18.68 & 33.82 & 02.35 & 06.62 \\
DeepSeek-R1      & \textbf{89.12} & \textbf{92.65} & \textbf{65.15} & \textbf{86.03} & \textbf{70.29} & \textbf{87.50} & \textbf{62.21} & \textbf{80.88} & \textbf{40.74} & \textbf{70.59} & \textbf{19.26} & \textbf{52.21} \\
Gemini-2.0-Flash & 83.82 & 86.76 & 19.12 & 26.47 & 30.00 & 43.38 & 22.79 & 32.35 & 19.26 & 28.68 & 00.00 & 00.00 \\
Gemini-2.5-Flash & 87.65 & 91.18 & 62.21 & \textbf{86.03} & 60.29 & 86.03 & 46.62 & 72.06 & 08.82 & 20.59 & 00.15 & 00.74 \\
GPT-3.5-Turbo    & 65.29 & 70.59 & 04.12 & 07.35 & 06.76 & 12.50 & 03.97 & 07.35 & 03.53 & 06.62 & 00.00 & 00.00 \\
GPT-4o           & 81.91 & 84.56 & 27.35 & 41.91 & 34.71 & 47.06 & 27.65 & 38.97 & 08.97 & 21.32 & 00.00 & 00.00 \\
GPT-4o-mini      & 75.88 & 79.41 & 06.18 & 11.76 & 10.88 & 19.85 & 05.88 & 13.97 & 01.91 & 05.15 & 00.00 & 00.00 \\
Grok-3           & 82.35 & 85.29 & 22.35 & 41.91 & 28.82 & 49.26 & 21.62 & 41.91 & 13.97 & 30.15 & 02.79 & 10.29 \\
Grok-3-Beta      & 82.35 & 89.71 & 26.47 & 48.53 & 28.53 & 54.41 & 21.76 & 43.38 & 16.18 & 31.62 & 02.06 & 06.62 \\
Mistral-Large    & 77.35 & 79.41 & 06.18 & 15.44 & 06.62 & 13.24 & 04.85 & 11.76 & 02.94 & 06.62 & 00.74 & 02.21 \\
Qwen-Coder-Plus  & 82.94 & 83.82 & 05.44 & 08.09 & 07.06 & 10.29 & 05.15 & 10.29 & 02.79 & 03.68 & 00.88 & 01.47 \\
Qwen-Coder-Turbo & 65.74 & 69.85 & 01.76 & 03.68 & 04.71 & 07.35 & 02.21 & 03.68 & 00.88 & 01.47 & 00.00 & 00.00 \\
Qwen2.5-Max         & 83.38 & 86.03 & 21.76 & 33.82 & 25.44 & 35.29 & 21.47 & 30.88 & 11.47 & 20.59 & 02.94 & 06.62 \\
Qwen2.5-Plus        & 83.09 & 86.03 & 12.65 & 25.00 & 14.56 & 30.88 & 05.15 & 15.44 & 01.62 & 05.88 & 00.00 & 00.00 \\
Qwen2.5-Turbo       & 78.53 & 81.62 & 01.91 & 03.68 & 03.68 & 06.62 & 01.47 & 05.15 & 00.44 & 00.73 & 00.00 & 00.00 \\
\bottomrule
\end{tabular}
\end{table*}

\section{Evaluation}\label{sec:evaluation}

%\subsection{Evaluation Setup}
We present our evaluation setups as follows before the discussion of research questions and evaluation results.

\textbf{Evaluated LLMs.} Our experiments evaluate \llmnum{} popular LLMs on \bench{}, generating SIMD-intrinsic code from natural language requirements.
The information of the models that are evaluated on \bench{}, including model names, tags, organizations, and sizes, is shown in Table~\ref{table:models}.
The selected \llmnum{} models are developed by seven leading organizations: Anthropic, Mistral AI, DeepSeek, Google, OpenAI, xAI, and Alibaba, which develop advanced LLMs that achieve outstanding results in scalar-code generation~\cite{EvalPlus-Leaderboard}.
For each organization, we select the most advanced LLMs (until our experiments) that our funding allows us to support.
LLMs that are representative but low-performing in scalar-code generation are excluded from our experiments, as the tasks in \bench{} are more challenging than scalar-code generation.
Among the selected LLMs, 3 are open-source and 15 are closed-source.

\textbf{Model settings.}
All evaluated LLMs are accessed by official APIs on the corresponding platforms.
We set the sampling temperature $T = 0.2$, which is suitable for code tasks.
For each task in \bench{}, we generate $N$ ($N = 5$) random samples and compute pass@1, pass@5, efficient@1, and efficient@5.
Although it is encouraged to generate more ($N \geq 5$) samples to avoid bias, we set $N = 5$ due to budget constraints.
Other model settings are default, and DeepSeek-R1 is the only reasoning model.

\textbf{Correctness evaluation environment.}
For convenience, we conduct correctness evaluations on the same machine rather than multiple machines on multiple architectures.
We conduct the correctness evaluation on a Docker container running Ubuntu 24.04.1 LTS in a Linux server, which is equipped with two AMD EPYC 7H12 64-Core CPUs (in the x86\_64 architecture), and each CPU has 512GB of RAM.
For scalar, SSE, and AVX implementations, programs can be compiled and executed directly, and we \texttt{clang} 18.1.3 for compilation. 
For Neon, SVE, and RVV implementations, we use cross-compilers (\texttt{clang} 19.1.7 for \texttt{aarch64} and \texttt{riscv64}) to generate executable files and QEMU (version 10.0.0) to run the executable files.
The timeout is set to 15.0 seconds for correctness evaluations.
For each correctness test case, the number of iterations is set to 1,000 during correctness evaluation.
%The details of the CPUs and compilers are shown in Table~\ref{table:env}.

\begin{comment}
\begin{table}[t]
\centering
\small
\caption{ CPUs and Compilers for Correctness Evaluations and Performance Evaluations.} \label{table:env}
\begin{tabular}{llll}
\toprule
\textbf{Eval} & \textbf{Intrinsic} & \textbf{CPU (Architecture)} & \textbf{Compiler} \\
\midrule
\multirow{5}{*}{\rotatebox{90}{\footnotesize \textbf{Correctness}}} 
& SSE       & AMD EPYC 7H12 64-Core (x86\_64) & clang 18.1.3 \\
& AVX       & AMD EPYC 7H12 64-Core (x86\_64) & clang 18.1.3 \\
& Neon      & AMD EPYC 7H12 64-Core (x86\_64) & clang 19.1.7 \\
& SVE       & AMD EPYC 7H12 64-Core (x86\_64) & clang 19.1.7 \\
& RVV       & AMD EPYC 7H12 64-Core (x86\_64) & clang 19.1.7 \\ \hline
\multirow{5}{*}{\rotatebox{90}{\footnotesize \textbf{Performance}}} 
& SSE & AMD EPYC 7H12 64-Core (x86\_64) & clang 18.1.3 \\
& AVX & AMD EPYC 7H12 64-Core (x86\_64) & clang 18.1.3 \\
& Neon & AWS EC2 C6g.8xlarge (AArch64) & clang 18.1.8 \\
& SVE & AWS EC2 C8g.8xlarge (AArch64)  & clang 18.1.8 \\
& RVV & -  & - \\
\bottomrule
\end{tabular}
\end{table}
\end{comment}

\textbf{Performance evaluation environment.}
Since the binary translation techniques used by emulators such as QEMU prioritize correctness, a SIMD instruction in the host processor is typically translated into multiple regular instructions in the guest processor to ensure correctness, leading to a significant performance loss~\cite{date2015qemu}.
For this reason, we conduct performance evaluations on multiple machines corresponding to different architectures to avoid the performance loss by binary translation.
For SSE and AVX implementations, the environment is the same as the environment of correctness evaluations.
For the Neon and SVE implementations, we use Amazon EC2 C6g (Graviton2) and C8g (Graviton4) instances in the AArch64 architecture, both with the 8xlarge size, the Amazon Linux operating system, and the \texttt{clang} compiler in version 18.1.8, which are provided by Amazon Web Services, Inc.
RVV-intrinsic experiments are excluded from our performance evaluations, as we find that few LLMs in our study can generate RVV-intrinsic programs that can pass the correctness test cases (as shown in Table~\ref{table:pass@k_results}).
The timeout is set to 150.0 seconds for performance evaluations.
Furthermore, we use commit \texttt{b20cea6} of the Google Benchmark library for our performance evaluations.

\textbf{Intrinsics and compiler options.}
Specific compiler options are required to enable the corresponding SIMD intrinsics. We list the selected intrinsics and the corresponding options in our experiments as follows.
(1) SSE/SSE2 intrinsics: \texttt{-msse -msse2}, 
(2) AVX/AVX2 intrinsics: \texttt{-mavx -mavx2},
(3) Neon intrinsics: \texttt{-march=armv8-a+simd},
(4) SVE intrinsics: \texttt{-march=armv8-a+sve},
and (5) RVV intrinsics: \texttt{-march=rv64gcv -mabi=lp64d -static}.
For SSE, AVX, and SVE intrinsics, multiple versions exist due to the continuous development of intrinsics. 
In our experiments, only the aforementioned basic and complete versions are selected, covering the processes of model output generation, compilation, and execution.
In correctness evaluations, the compiler optimization option \texttt{-O0} is used, and both \texttt{-O0} and \texttt{-O3} options are used in performance evaluations.

\subsection{RQ1: Correctness Evaluation}

In this section, we answer \textbf{RQ1} (\textit{How do LLMs perform in generating SIMD-intrinsic code that is semantically correct?}) by measuring pass@k across six scenarios of code generation: scalar, SSE, AVX, Neon, SVE, and RVV.
The detailed results of this section are presented in Table~\ref{table:pass@k_results}.
We also generate scalar samples for each task in \bench{} as a baseline. We discuss how to reuse the prompts and test cases for scalar-code generation in Section~\ref{sec:task-construction}.

\textbf{Scalar code generation vs. SIMD-intrinsic code generation.}
As shown in Table~\ref{table:pass@k_results}, almost all evaluated LLMs achieve high pass@k results in generating scalar code for tasks of \bench{}, being consistent with the results in EvalPlus~\cite{EvalPlus-Leaderboard} (a representative benchmark for scalar-code generation based on HumanEval~\cite{humaneval2021}).
All models except GPT-3.5-Turbo and Qwen-Coder-Turbo achieve a pass@1 rate exceeding 75\% when generating scalar code.
However, all LLMs exhibit a decrease in pass@k for SIMD-intrinsic code generation, highlighting the significant limitations of current LLMs in handling these challenging scenarios.
The decrease in pass@k is particularly noticeable for Codestral, GPT-3.5-Turbo, GPT-4o-mini, Mistral-Large, Qwen-Coder-Plus, Qwen-Coder-Turbo, and Qwen2.5-Turbo. These seven LLMs exhibit pass@1 $\leq 15\%$ and pass@5 $\leq 20\%$ in all five intrinsic types, although the pass@k results of these models for scalar-code generation are commendable.

\textbf{Comparison among LLMs.}
In the correctness evaluations, DeepSeek-R1 achieves the best pass@k results among all models, resulting in the highest pass@1 and pass@5 in all six scenarios (scalar, SSE, AVX, Neon, SVE, and RVV).
The results of DeepSeek-R1 are especially outstanding in the scenarios of SVE and RVV (the only two intrinsic types with scalable vector lengths and VLA programming models, as discussed in Section~\ref{sec2:mainsimd}), while other evaluated LLMs rarely generate semantically correct code in these two scenarios.
Overall, Gemini-2.5-Flash achieves pass@k results second to only DeepSeek-R1.
Gemini-2.5-Flash shares the highest pass@5 for SSE with DeepSeek-R1 and achieves the second highest pass@k in the scenarios of scalar, AVX, and Neon.

\textbf{Comparison among intrinsic types.}
Among the five evaluated intrinsics, 17 of 18 LLMs (with Mistral-Large as the only exception) achieve the highest pass@5 on AVX intrinsics, while all 18 LLMs exhibit their lowest pass@5 on RVV intrinsics.
Overall, LLMs exhibit higher correctness of SIMD-intrinsic code generation for SSE, AVX, and Neon than for SVE and RVV.
Among the 18 evaluated LLMs, 17 (with Gemini-2.0-Flash as the only exception) exhibit the two lowest pass@5 rankings for SVE and RVV intrinsics.
LLMs exhibit their lowest rankings in the RVV-intrinsic scenario, with seven LLMs failing to generate any correct compilations across all tasks in \bench{}.
We discuss the detailed error types of the compilations for each intrinsic in Section~\ref{sec:4-RQ3} and the potential solutions of these errors in Section~\ref{sec:discussion}.

\subsection{RQ2: Performance Evaluation}

\begin{figure*}[t]
\includegraphics[width=0.975\textwidth]{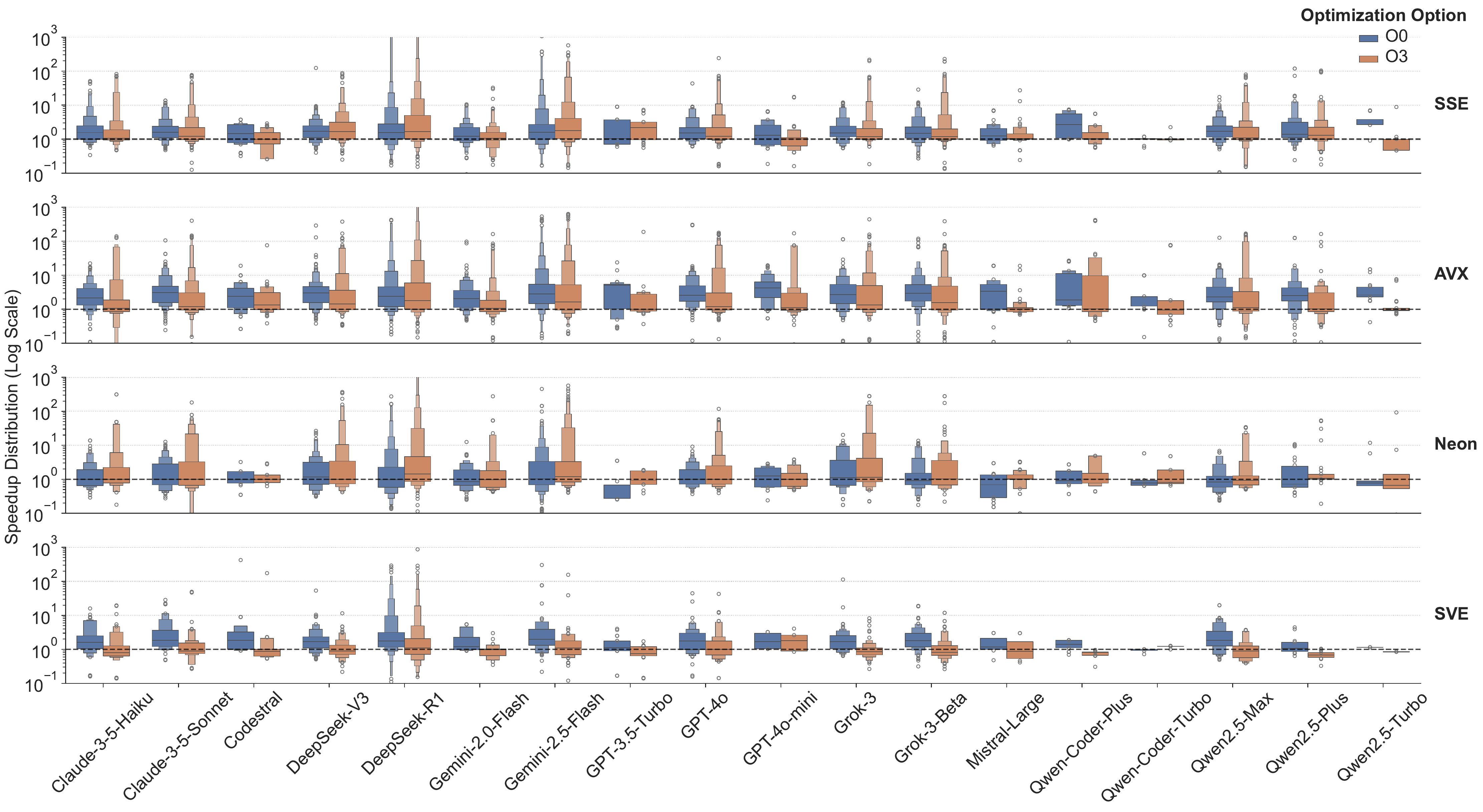}
\centering
\caption{Distribution of Speedup Results for Valid Samples.}
\label{fig:performance}
\Description[<short description>]{<long description>}
\end{figure*}

\begin{figure}[t]
\includegraphics[width=0.475\textwidth]{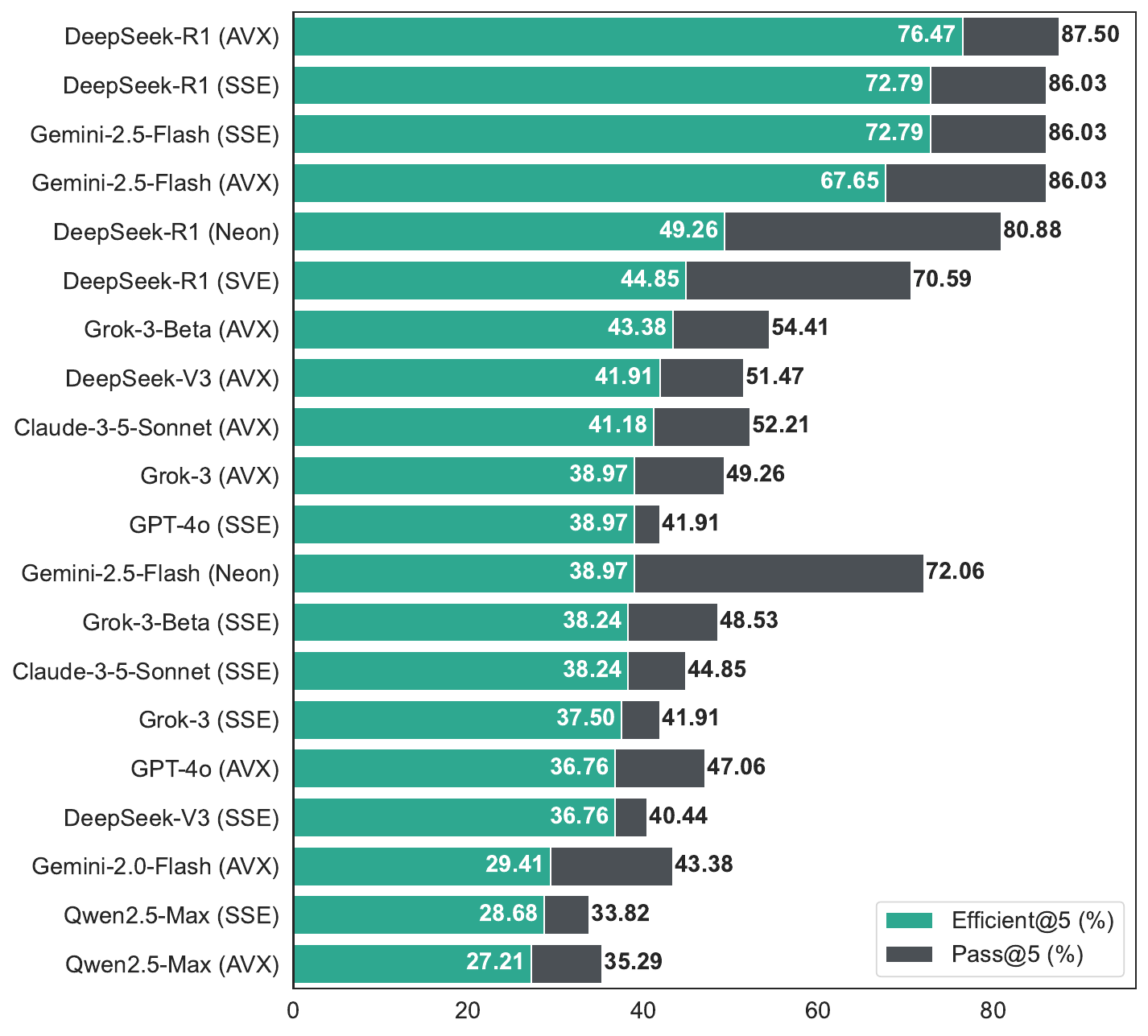}
\centering
\caption{Top-20 Efficient@5 and Pass@5 Results at \texttt{-O3}, Ranked by Efficient@5 (\%).}
\label{fig:performance-rank}
\Description[<short description>]{<long description>}
\end{figure}

In this section, we answer \textbf{RQ2} (\textit{What is the performance of the valid SIMD-intrinsic code generated by LLMs?}) by measuring the speedup results against scalar implementations using the performance test cases from \bench{} across four scenarios of code generation: SSE, AVX, Neon, and SVE.
As discussed at the beginning of Section~\ref{sec:evaluation}, RVV-intrinsic experiments are excluded from our performance evaluations, as we find that few LLMs in our study can generate valid RVV-intrinsic programs.
The main results of the performance evaluations are presented in Figure~\ref{fig:performance} and Figure~\ref{fig:performance-rank}.
Figure~\ref{fig:performance} shows the distribution of the speedup results for only valid samples that can pass the correctness tests, as evaluating the performance of invalid code, which leads to crashes or incorrect results, is meaningless.
Figure~\ref{fig:performance-rank} presents the ranking of efficient@5 results at \texttt{-O3} for comparing the performance results across LLMs and intrinsics.

We first discuss how the speedup in Figure~\ref{fig:performance} is derived from the code samples.
Given an optimization option and a code sample for a specific intrinsic generated by an LLM, the canonical solution in scalar, the vectorized code sample, and the corresponding performance test case (implemented using Google Benchmark) are compiled by the same compiler using the same option and executed on the same machine.
Thus, the only variable in performance evaluations is code implementation (scalar or vectorized), while other factors, such as compiler, optimization option, and machine architecture, remain constant.
The details of how performance test cases are implemented and how speedup is calculated to avoid bias are discussed in Section~\ref{sec:test-construction}.

\textbf{Compiler optimization vs. SIMD-intrinsic vectorization.}
As shown in Figure~\ref{fig:performance}, we find that the valid SIMD-intrinsic code can achieve speed $>1.0$ for a substantial number of cases at both \texttt{-O0} and \texttt{-O3}.
This finding indicates that LLM-assisted vectorized-code programming with SIMD intrinsics can effectively fill the limitations of compiler auto-vectorization and other optimizations, resulting in a higher peak performance than scalar-code programming.
The performance improvement when compiling at \texttt{-O3} stems from two factors.
(1) Explicit vectorization via SIMD intrinsics enables more precise generation of SIMD instructions for fine-grained performance optimization than compiler auto-vectorization.
(2) Compiler optimizations (theoretically excluding auto-vectorization optimizations) can enable additional optimizations based on the SIMD instructions generated from SIMD intrinsics, resulting in further performance improvements.

Note that LLM-assisted vectorized code programming with SIMD intrinsics does not always outperform compiler optimizations; the former's effectiveness depends on the capabilities of the LLMs and the optimization potential of the problems.
For example, among the 24 valid AVX samples generated by Qwen2.5-Turbo, 95.83\% (23/24) samples exhibit speedup $>1.0$ at \texttt{-O0}; however, only 41.67\% (10/24) samples exhibit speedup $>1.0$ at \texttt{-O3}.

\textbf{Comparison among LLMs and intrinsic types.}
Figure~\ref{fig:performance-rank} presents the ranking of the efficient@5 results.
We find that a high efficient@5 is typically associated with a high pass@5.
The LLMs that perform well in pass@5, such as DeepSeek-R1 and Gemini-2.5-Flash, also exhibit strong effectiveness in efficient@5.
A notable exception is Gemini-2.5-Flash for Neon intrinsics, which exhibits a pass@5 of 72.06, but an efficient@5 of only 38.97.
We also find that LLMs perform better during performance evaluation for SSE and AVX than for Neon and SVE.
Among the top-20 results in Figure~\ref{fig:performance-rank}, 17 (85\%) are for SSE and AVX, 2 are for Neon, and only 1 is for SVE.
%Improving the cross-platform portability of the code generated by LLMs is essential.
%\textbf{Case study.}

\subsection{RQ3: Invalid Case Analysis}\label{sec:4-RQ3}

\begin{figure}[t]
\includegraphics[width=0.475\textwidth]{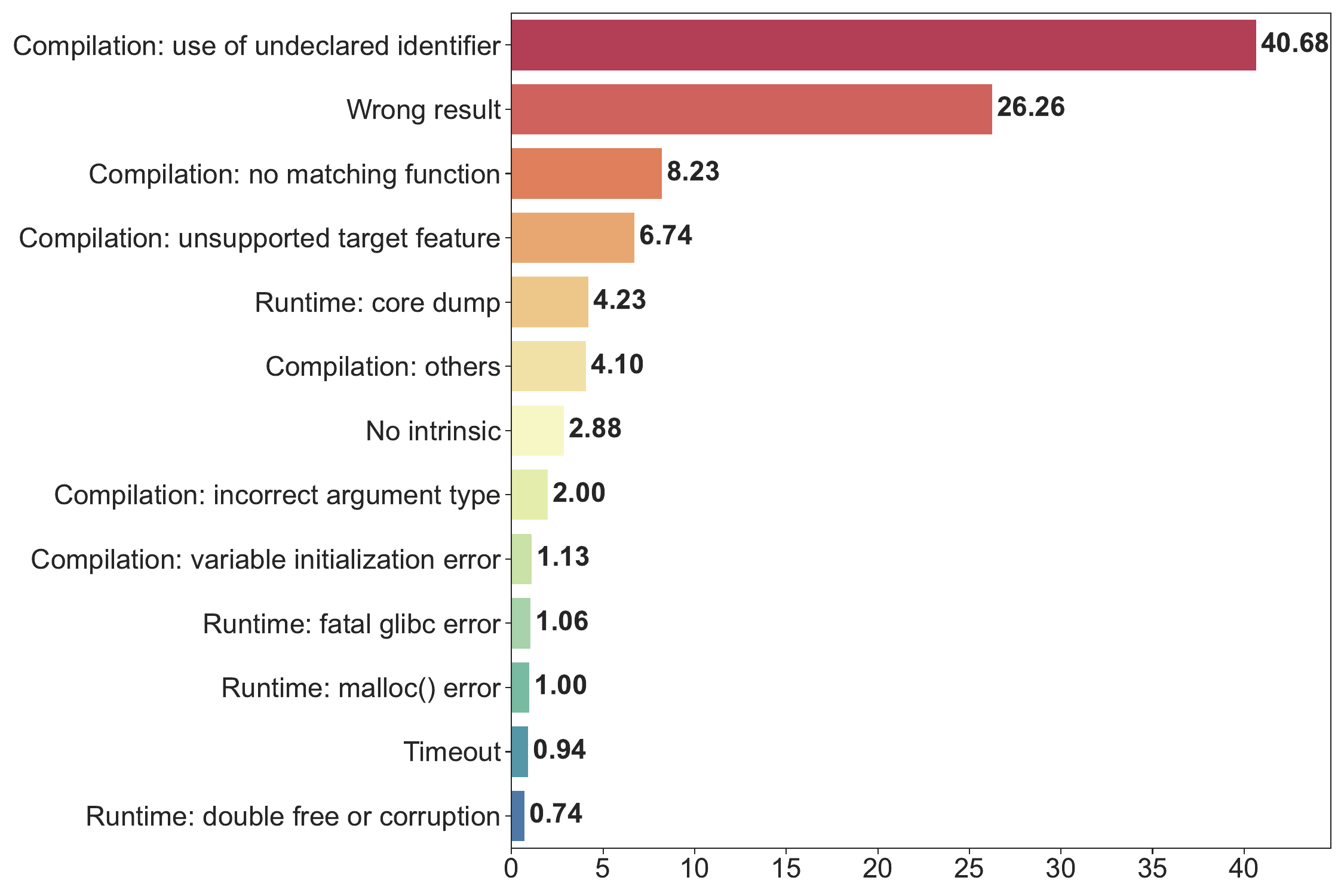}
\centering
\caption{Distribution of Reasons for Invalid Cases (\%).}
\label{fig:error-dist}
\Description[<short description>]{<long description>}
\end{figure}

\begin{table*}[t]
    \centering
    \caption{Top-3 Error Types for Each Intrinsic Type.}
    \label{tab:top3-error}
\begin{threeparttable}
    \begin{tabular}{l|r|r|r}
    \toprule
    \textbf{Intrinsic} & \textbf{1st} & \textbf{2nd} & \textbf{3rd}  \\ \midrule
    SSE & Wrong result (36.32\%) & Compilation-target$^{2}$ (29.92\%) & Compilation-identifier$^{1}$ (12.26\%) \\
    AVX & Wrong result (47.41\%) & Compilation-identifier$^{1}$ (18.16\%) & Runtime: core dump (7.64\%) \\
    Neon& Wrong result (41.47\%) & Compilation-identifier$^{1}$ (20.14\%) & Runtime: core dump (7.79\%) \\
    SVE & Compilation-identifier$^{1}$ (51.11\%) & Compilation-function$^{3}$ (22.72\%) & Wrong result (11.00\%) \\
    RVV & Compilation-identifier$^{1}$ (88.91\%) & Compilation-function$^{3}$ (3.82\%) & Wrong result (2.99\%) \\ 
    \bottomrule
    \end{tabular}
\begin{tablenotes}
\item[1] Compilation-identifier: errors caused by the use of undeclared identifiers.
\item[2] Compilation-target: errors caused by unsupported target features.
\item[3] Compilation-function: errors caused by no matching function.
\end{tablenotes}
\end{threeparttable}
\end{table*}

In this section, we answer \textbf{RQ3} (\textit{What are the common error types in SIMD-intrinsic code generated by LLMs?}) by categorizing the compilation and execution error messages.

The error messages are categorized into five main types. 
(1) Compilation error. Programs of this type cannot be successfully compiled due to reasons such as syntax errors and type errors.
(2) Runtime error. Programs of this type can be successfully compiled but crash during execution due to reasons such as segment faults.
(3) Wrong result. Programs of this type can be successfully compiled and executed, but produce incorrect results due to logical bugs (i.e., cannot pass the correctness test cases).
(4) No intrinsic. LLMs fail to adhere to the requirements in the prompts for using SIMD intrinsics to enable parallelism, instead generating scalar code, regardless of the correctness of the generated code.
(5) Timeout. Programs of this type cannot be compiled or executed within a given time.

Figure~\ref{fig:error-dist} shows the distribution of error types for invalid cases in our experiments.
The two primary reasons for invalid cases are compilation errors (caused by the use of undeclared identifiers) and producing incorrect results, which account for 40.68\% and 26.26\%, respectively.
The compilation errors involving ``use of undeclared identifier'' are primarily caused by the generation of programs with incorrect intrinsic invocations, where the innovated intrinsics have similar but distinct names compared to the correct ones.
The wrong results (i.e., logical bugs) are common in the task of code generation.
As we discussed in Section~\ref{sec:intro}, programming vectorized code using SIMD intrinsics is more challenging than programming conventional scalar code, as extra operations such as data alignment are required to ensure correctness.
The challenges inherent to the tasks result in a decrease in the correctness of vectorized code generated by LLMs, compared to scalar-code generation.
The compilation errors involving ``no matching function'' rank third (8.23\%).
Compilation errors of this type can occur when the function's parameters (e.g., overloaded intrinsics) do not match the number, type, or constness of the arguments.

We also analyze the top-3 error types for each intrinsic type, and the results are shown in Table~\ref{tab:top3-error}, leading to another three interesting findings.
(1) For SSE, AVX, and Neon intrinsics, LLMs struggle mainly against logical bugs, while for SVE and RVV intrinsics, LLMs struggle mainly against compilation errors (especially compilation errors involving ``use of undeclared identifier'').
(2) Compilation errors involving ``unsupported target feature'' are commonly observed during the evaluation of SSE intrinsics.
We limit the SSE versions under evaluation to the basic versions (SSE and SSE2) as discussed at the beginning of Section~\ref{sec:evaluation}, while some LLMs ignore this limitation and generate intrinsics from advanced SSE versions, leading to errors of this type.
(3) Almost all invalid cases of RVV intrinsics are caused by compilation errors involving ``use of undeclared identifier'' and ``no matching function''.
The most common issue is the absence of the intrinsic prefix \texttt{\_\_riscv\_}, which is introduced in an update to the RVV intrinsics documentation~\cite{RVV-intrinsic}.
We hypothesize that the underlying cause is that the training data of the LLMs is neither updated with new corpora of RVV intrinsics nor purged of outdated data.

\section{Discussion}\label{sec:discussion}

From the preceding experiments, we find that SIMD-intrinsic code generation by LLMs still faces multiple limitations.
In this section, we discuss promising directions for the further advancement of LLMs in this task, as well as potential application scenarios.

We identify three promising directions for advancing LLMs in SIMD-intrinsic code generation.
(1) Develop high-quality training datasets that include up-to-date SIMD intrinsics.
As discussed in Section~\ref{sec:4-RQ3}, current LLMs struggle with generating SVE and RVV code due to frequent compilation errors such as ``use of undeclared identifier'', primarily caused by the absence of relevant up-to-date intrinsic definitions in existing training data.
%LLMs face major limitations in SIMD-intrinsic code generation for SVE and RVV mainly due to compilation errors involving ``use of undeclared identifier'', which are mainly caused by the absence of data entries with the latest SIMD intrinsics.
(2) Incorporate retrieval-augmented generation (RAG) during code generation.
%The RAG approach can retrieve relevant information (e.g., the definitions of potentially used intrinsics) from the external SIMD documents to avoid errors such as undeclared identifiers.
RAG enables the LLMs to access external SIMD documentation to retrieve relevant information (such as intrinsic definitions), thereby reducing errors caused by incorrect or unknown identifiers.
(3) Adopt a step-by-step generation strategy, where LLMs first generate scalar code and then apply vectorization using SIMD intrinsics. 
This process can be further enhanced by transformation techniques such as those proposed in~\cite{zheng2025vectrans}, which can improve the performance and correctness of the generated vectorized code.

LLM-based SIMD-intrinsic code generation holds significant potential for optimizing software with hardware-level features.
(1) LLMs can assist developers in optimizing existing performance-critical libraries, as discussed in Section~\ref{sec:intro}.
(2) LLMs can improve the cross-platform portability of SIMD-intrinsic code, which is currently limited by architecture-specific intrinsic differences; LLMs offer a promising solution to bridge these gaps.
%SIMD-intrinsic code is currently non-portable due to the differences in intrinsics across architectures, but this limitation could be alleviated with the help of LLMs.
(3) LLMs can enhance the security and reliability of toolchains for SIMD intrinsics.
While fuzzing tools~\cite{he2025rvismithfuzzingcompilersrvv} have been proposed to test SIMD-intrinsic compilers, LLMs can support more extensive and systematic testing.

\section{Conclusion}

In this paper, we have proposed \bench{}, the first code benchmark specially designed for SIMD-intrinsic code generation, comprising \tasknum{} carefully crafted tasks and targeting five representative SIMD intrinsics: SSE, AVX, Neon, SVE, and RVV.
We have conducted a systematic evaluation (including both correctness and performance) for \llmnum{} representative LLMs on \bench{}, resulting in a series of novel and insightful findings.
Our evaluation results have demonstrated that LLMs exhibit a universal decrease in SIMD-intrinsic code generation compared to scalar-code generation, and the valid code generated by LLMs using SIMD intrinsics results in further performance improvements compared to scalar code in a significant number of cases.
The main current obstacles to SIMD-intrinsic code generation have been demonstrated as compilation errors related to the ``use of undeclared identifier'' and logical bugs in the generated code.
Our in-depth analysis has highlighted promising directions for the further advancement of LLMs in the challenging scenarios of SIMD-intrinsic code generation.

\begin{acks}
This work was partially supported by National Natural Science Foundation of China under Grant No. 92464301 and Damo Academy (Hupan Laboratory) through Damo Academy (Hupan Laboratory) Innovative Research Program. 
\end{acks}

\balance
\bibliographystyle{ACM-Reference-Format}
\bibliography{root}


\begin{thebibliography}{58}

%%% the Web address.

\ifx \showCODEN    \undefined \def \showCODEN     #1{\unskip}     \fi
\ifx \showISBNx    \undefined \def \showISBNx     #1{\unskip}     \fi
\ifx \showISBNxiii \undefined \def \showISBNxiii  #1{\unskip}     \fi
\ifx \showISSN     \undefined \def \showISSN      #1{\unskip}     \fi
\ifx \showLCCN     \undefined \def \showLCCN      #1{\unskip}     \fi
\ifx \shownote     \undefined \def \shownote      #1{#1}          \fi
\ifx \showarticletitle \undefined \def \showarticletitle #1{#1}   \fi
\ifx \showURL      \undefined \def \showURL       {\relax}        \fi
% invisible to TeX
\providecommand\bibfield[2]{#2}
\providecommand\bibinfo[2]{#2}
\providecommand\natexlab[1]{#1}
\providecommand\showeprint[2][]{arXiv:#2}

\bibitem[AI(2024)]%
        {Mistral-Large}
\bibfield{author}{\bibinfo{person}{Mistral AI}.} \bibinfo{year}{2024}\natexlab{}.
\newblock \bibinfo{title}{Mistral-Large-Instruct-2411}.
\newblock \bibinfo{howpublished}{\url{https://huggingface.co/mistralai/Mistral-Large-Instruct-2411}}.
\newblock


\bibitem[AI(2025)]%
        {Codestral}
\bibfield{author}{\bibinfo{person}{Mistral AI}.} \bibinfo{year}{2025}\natexlab{}.
\newblock \bibinfo{title}{Codestral 25.01}.
\newblock \bibinfo{howpublished}{\url{https://mistral.ai/news/codestral-2501}}.
\newblock


\bibitem[Algorithmica(2025)]%
        {masking}
\bibfield{author}{\bibinfo{person}{Algorithmica}.} \bibinfo{year}{2025}\natexlab{}.
\newblock \bibinfo{title}{Masking and Blending}.
\newblock \bibinfo{howpublished}{\url{https://en.algorithmica.org/hpc/simd/masking/}}.
\newblock


\bibitem[Anthropic(2025)]%
        {claude}
\bibfield{author}{\bibinfo{person}{Anthropic}.} \bibinfo{year}{2025}\natexlab{}.
\newblock \bibinfo{title}{Claude}.
\newblock \bibinfo{howpublished}{\url{https://www.anthropic.com/claude/}}.
\newblock


\bibitem[ARM(2023)]%
        {Neon}
\bibfield{author}{\bibinfo{person}{ARM}.} \bibinfo{year}{2023}\natexlab{}.
\newblock \bibinfo{title}{Neon}.
\newblock \bibinfo{howpublished}{\url{https://developer.arm.com/Architectures/Neon}}.
\newblock


\bibitem[ARM(2025)]%
        {SVE}
\bibfield{author}{\bibinfo{person}{ARM}.} \bibinfo{year}{2025}\natexlab{}.
\newblock \bibinfo{title}{SVE Optimization Guide}.
\newblock \bibinfo{howpublished}{\url{https://developer.arm.com/documentation/102699/0100}}.
\newblock


\bibitem[Athiwaratkun et~al\mbox{.}(2023)]%
        {multilingual2023}
\bibfield{author}{\bibinfo{person}{Ben Athiwaratkun}, \bibinfo{person}{Sanjay~Krishna Gouda}, \bibinfo{person}{Zijian Wang}, \bibinfo{person}{Xiaopeng Li}, \bibinfo{person}{Yuchen Tian}, \bibinfo{person}{Ming Tan}, \bibinfo{person}{Wasi~Uddin Ahmad}, \bibinfo{person}{Shiqi Wang}, \bibinfo{person}{Qing Sun}, \bibinfo{person}{Mingyue Shang}, \bibinfo{person}{Sujan~Kumar Gonugondla}, \bibinfo{person}{Hantian Ding}, \bibinfo{person}{Varun Kumar}, \bibinfo{person}{Nathan Fulton}, \bibinfo{person}{Arash Farahani}, \bibinfo{person}{Siddhartha Jain}, \bibinfo{person}{Robert Giaquinto}, \bibinfo{person}{Haifeng Qian}, \bibinfo{person}{Murali~Krishna Ramanathan}, \bibinfo{person}{Ramesh Nallapati}, \bibinfo{person}{Baishakhi Ray}, \bibinfo{person}{Parminder Bhatia}, \bibinfo{person}{Sudipta Sengupta}, \bibinfo{person}{Dan Roth}, {and} \bibinfo{person}{Bing Xiang}.} \bibinfo{year}{2023}\natexlab{}.
\newblock \bibinfo{title}{Multi-lingual Evaluation of Code Generation Models}.
\newblock
\showeprint[arxiv]{2210.14868}~[cs.LG]
\urldef\tempurl%
\url{https://arxiv.org/abs/2210.14868}
\showURL{%
\tempurl}


\bibitem[Austin et~al\mbox{.}(2021)]%
        {austin2021program}
\bibfield{author}{\bibinfo{person}{Jacob Austin}, \bibinfo{person}{Augustus Odena}, \bibinfo{person}{Maxwell Nye}, \bibinfo{person}{Maarten Bosma}, \bibinfo{person}{Henryk Michalewski}, \bibinfo{person}{David Dohan}, \bibinfo{person}{Ellen Jiang}, \bibinfo{person}{Carrie Cai}, \bibinfo{person}{Michael Terry}, \bibinfo{person}{Quoc Le}, {and} \bibinfo{person}{Charles Sutton}.} \bibinfo{year}{2021}\natexlab{}.
\newblock \bibinfo{title}{Program Synthesis with Large Language Models}.
\newblock
\showeprint[arxiv]{2108.07732}~[cs.PL]
\urldef\tempurl%
\url{https://arxiv.org/abs/2108.07732}
\showURL{%
\tempurl}


\bibitem[Author(s)(2025)]%
        {simdbench}
\bibfield{author}{\bibinfo{person}{Anonymous Author(s)}.} \bibinfo{year}{2025}\natexlab{}.
\newblock \bibinfo{title}{SimdBench}.
\newblock \bibinfo{howpublished}{\url{https://anonymous.4open.science/r/SimdBench-1B3F/}}.
\newblock


\bibitem[Baghsorkhi et~al\mbox{.}(2016)]%
        {vectorization-pldi16}
\bibfield{author}{\bibinfo{person}{Sara~S. Baghsorkhi}, \bibinfo{person}{Nalini Vasudevan}, {and} \bibinfo{person}{Youfeng Wu}.} \bibinfo{year}{2016}\natexlab{}.
\newblock \showarticletitle{FlexVec: Auto-Vectorization for Irregular Loops}. In \bibinfo{booktitle}{\emph{Proceedings of the 37th ACM SIGPLAN Conference on Programming Language Design and Implementation}}.
\newblock
\href{https://doi.org/10.1145/2980983.2908111}{doi:\nolinkurl{10.1145/2980983.2908111}}


\bibitem[Chen et~al\mbox{.}(2021b)]%
        {humaneval2021}
\bibfield{author}{\bibinfo{person}{Mark Chen}, \bibinfo{person}{Jerry Tworek}, \bibinfo{person}{Heewoo Jun}, \bibinfo{person}{Qiming Yuan}, \bibinfo{person}{Henrique Ponde De~Oliveira Pinto}, \bibinfo{person}{Jared Kaplan}, \bibinfo{person}{Harri Edwards}, \bibinfo{person}{Yuri Burda}, \bibinfo{person}{Nicholas Joseph}, \bibinfo{person}{Greg Brockman}, {et~al\mbox{.}}} \bibinfo{year}{2021}\natexlab{b}.
\newblock \bibinfo{title}{Evaluating Large Language Models Trained on Code}.
\newblock
\showeprint[arxiv]{2107.03374}~[cs.LG]
\urldef\tempurl%
\url{https://arxiv.org/abs/2107.03374}
\showURL{%
\tempurl}


\bibitem[Chen et~al\mbox{.}(2021a)]%
        {vectorization-asplos21}
\bibfield{author}{\bibinfo{person}{Yishen Chen}, \bibinfo{person}{Charith Mendis}, \bibinfo{person}{Michael Carbin}, {and} \bibinfo{person}{Saman Amarasinghe}.} \bibinfo{year}{2021}\natexlab{a}.
\newblock \showarticletitle{VeGen: A Vectorizer Generator for SIMD and Beyond}. In \bibinfo{booktitle}{\emph{Proceedings of the 26th ACM International Conference on Architectural Support for Programming Languages and Operating Systems}}.
\newblock
\href{https://doi.org/10.1145/3445814.3446692}{doi:\nolinkurl{10.1145/3445814.3446692}}


\bibitem[DeepSeek-AI(2024)]%
        {deepseekai2024deepseekv3technicalreport}
\bibfield{author}{\bibinfo{person}{DeepSeek-AI}.} \bibinfo{year}{2024}\natexlab{}.
\newblock \bibinfo{title}{DeepSeek-V3 Technical Report}.
\newblock
\showeprint[arxiv]{2412.19437}~[cs.CL]
\urldef\tempurl%
\url{https://arxiv.org/abs/2412.19437}
\showURL{%
\tempurl}


\bibitem[DeepSeek-AI(2025)]%
        {deepseekai2025deepseekr1}
\bibfield{author}{\bibinfo{person}{DeepSeek-AI}.} \bibinfo{year}{2025}\natexlab{}.
\newblock \bibinfo{title}{DeepSeek-R1: Incentivizing Reasoning Capability in LLMs via Reinforcement Learning}.
\newblock
\showeprint[arxiv]{2501.12948}~[cs.CL]
\urldef\tempurl%
\url{https://arxiv.org/abs/2501.12948}
\showURL{%
\tempurl}


\bibitem[Du et~al\mbox{.}(2024b)]%
        {Mercury2024nips}
\bibfield{author}{\bibinfo{person}{Mingzhe Du}, \bibinfo{person}{Luu~Anh Tuan}, \bibinfo{person}{Bin Ji}, \bibinfo{person}{Qian Liu}, {and} \bibinfo{person}{See-Kiong Ng}.} \bibinfo{year}{2024}\natexlab{b}.
\newblock \showarticletitle{Mercury: A Code Efficiency Benchmark for Code Large Language Models}. In \bibinfo{booktitle}{\emph{Proceedings of the 38th Annual Conference on Neural Information Processing Systems}}.
\newblock
\urldef\tempurl%
\url{https://arxiv.org/abs/2402.07844}
\showURL{%
\tempurl}


\bibitem[Du et~al\mbox{.}(2024a)]%
        {classeval2023}
\bibfield{author}{\bibinfo{person}{Xueying Du}, \bibinfo{person}{Mingwei Liu}, \bibinfo{person}{Kaixin Wang}, \bibinfo{person}{Hanlin Wang}, \bibinfo{person}{Junwei Liu}, \bibinfo{person}{Yixuan Chen}, \bibinfo{person}{Jiayi Feng}, \bibinfo{person}{Chaofeng Sha}, \bibinfo{person}{Xin Peng}, {and} \bibinfo{person}{Yiling Lou}.} \bibinfo{year}{2024}\natexlab{a}.
\newblock \showarticletitle{Evaluating Large Language Models in Class-Level Code Generation}. In \bibinfo{booktitle}{\emph{Proceedings of the IEEE/ACM 46th International Conference on Software Engineering}}.
\newblock
\href{https://doi.org/10.1145/3597503.3639219}{doi:\nolinkurl{10.1145/3597503.3639219}}


\bibitem[Feng et~al\mbox{.}(2021)]%
        {limit_autov1}
\bibfield{author}{\bibinfo{person}{Jing~Ge Feng}, \bibinfo{person}{Ye~Ping He}, \bibinfo{person}{Qiu~Ming Tao}, {and} \bibinfo{person}{Fazli Wahid}.} \bibinfo{year}{2021}\natexlab{}.
\newblock \showarticletitle{Evaluation of Compilers’ Capability of Automatic Vectorization Based on Source Code Analysis}.
\newblock \bibinfo{journal}{\emph{Scientific Programming}} (\bibinfo{year}{2021}).
\newblock
\href{https://doi.org/10.1155/2021/3264624}{doi:\nolinkurl{10.1155/2021/3264624}}


\bibitem[Fu et~al\mbox{.}(2015)]%
        {date2015qemu}
\bibfield{author}{\bibinfo{person}{Sheng-Yu Fu}, \bibinfo{person}{Jan-Jan Wu}, {and} \bibinfo{person}{Wei-Chung Hsu}.} \bibinfo{year}{2015}\natexlab{}.
\newblock \showarticletitle{Improving SIMD Code Generation in QEMU}. In \bibinfo{booktitle}{\emph{Proceedings of the 2015 Design, Automation \& Test in Europe Conference \& Exhibition}}. \bibinfo{pages}{1233–1236}.
\newblock


\bibitem[Google(2025a)]%
        {gemini-api}
\bibfield{author}{\bibinfo{person}{Google}.} \bibinfo{year}{2025}\natexlab{a}.
\newblock \bibinfo{title}{Gemini Developer API}.
\newblock \bibinfo{howpublished}{\url{https://ai.google.dev/gemini-api/docs}}.
\newblock


\bibitem[Google(2025b)]%
        {googlebenchmark}
\bibfield{author}{\bibinfo{person}{Google}.} \bibinfo{year}{2025}\natexlab{b}.
\newblock \bibinfo{title}{A microbenchmark support library}.
\newblock \bibinfo{howpublished}{\url{https://github.com/google/benchmark}}.
\newblock


\bibitem[Google(2025c)]%
        {XNNPACK}
\bibfield{author}{\bibinfo{person}{Google}.} \bibinfo{year}{2025}\natexlab{c}.
\newblock \bibinfo{title}{XNNPACK}.
\newblock \bibinfo{howpublished}{\url{https://github.com/google/XNNPACK}}.
\newblock


\bibitem[Google(2025d)]%
        {geminiteam2025geminifamily}
\bibfield{author}{\bibinfo{person}{Gemini~Team Google}.} \bibinfo{year}{2025}\natexlab{d}.
\newblock \bibinfo{title}{Gemini: A Family of Highly Capable Multimodal Models}.
\newblock
\showeprint[arxiv]{2312.11805}~[cs.CL]
\urldef\tempurl%
\url{https://arxiv.org/abs/2312.11805}
\showURL{%
\tempurl}


\bibitem[He et~al\mbox{.}(2025)]%
        {he2025rvismithfuzzingcompilersrvv}
\bibfield{author}{\bibinfo{person}{Yibo He}, \bibinfo{person}{Cunjian Huang}, \bibinfo{person}{Xianmiao Qu}, \bibinfo{person}{Hongdeng Chen}, \bibinfo{person}{Wei Yang}, {and} \bibinfo{person}{Tao Xie}.} \bibinfo{year}{2025}\natexlab{}.
\newblock \bibinfo{title}{RVISmith: Fuzzing Compilers for RVV Intrinsics}.
\newblock
\showeprint[arxiv]{2507.03773}~[cs.CR]
\urldef\tempurl%
\url{https://arxiv.org/abs/2507.03773}
\showURL{%
\tempurl}


\bibitem[Hendrycks et~al\mbox{.}(2021)]%
        {apps2021}
\bibfield{author}{\bibinfo{person}{Dan Hendrycks}, \bibinfo{person}{Steven Basart}, \bibinfo{person}{Saurav Kadavath}, \bibinfo{person}{Mantas Mazeika}, \bibinfo{person}{Akul Arora}, \bibinfo{person}{Ethan Guo}, \bibinfo{person}{Collin Burns}, \bibinfo{person}{Samir Puranik}, \bibinfo{person}{Horace He}, \bibinfo{person}{Dawn Song}, {and} \bibinfo{person}{Jacob Steinhardt}.} \bibinfo{year}{2021}\natexlab{}.
\newblock \bibinfo{title}{Measuring Coding Challenge Competence With APPS}.
\newblock
\showeprint[arxiv]{2105.09938}~[cs.SE]
\urldef\tempurl%
\url{https://arxiv.org/abs/2105.09938}
\showURL{%
\tempurl}


\bibitem[hqztrue(2025)]%
        {leetcodesolution}
\bibfield{author}{\bibinfo{person}{hqztrue}.} \bibinfo{year}{2025}\natexlab{}.
\newblock \bibinfo{title}{GitHub Repository: LeetCodeSolutions}.
\newblock \bibinfo{howpublished}{\url{https://github.com/hqztrue/LeetCodeSolutions/blob/master/1801-1900/}}.
\newblock


\bibitem[Huang et~al\mbox{.}(2024)]%
        {EffiBench2024nips}
\bibfield{author}{\bibinfo{person}{Dong Huang}, \bibinfo{person}{Yuhao Qing}, \bibinfo{person}{Weiyi Shang}, \bibinfo{person}{Heming Cui}, {and} \bibinfo{person}{Jie~M. Zhang}.} \bibinfo{year}{2024}\natexlab{}.
\newblock \showarticletitle{EffiBench: Benchmarking the Efficiency of Automatically Generated Code}. In \bibinfo{booktitle}{\emph{Proceedings of the 38th Annual Conference on Neural Information Processing Systems}}.
\newblock
\urldef\tempurl%
\url{https://arxiv.org/abs/2402.02037}
\showURL{%
\tempurl}


\bibitem[Hui et~al\mbox{.}(2024)]%
        {hui2024qwen25codertechnicalreport}
\bibfield{author}{\bibinfo{person}{Binyuan Hui}, \bibinfo{person}{Jian Yang}, \bibinfo{person}{Zeyu Cui}, \bibinfo{person}{Jiaxi Yang}, \bibinfo{person}{Dayiheng Liu}, \bibinfo{person}{Lei Zhang}, \bibinfo{person}{Tianyu Liu}, \bibinfo{person}{Jiajun Zhang}, \bibinfo{person}{Bowen Yu}, \bibinfo{person}{Keming Lu}, \bibinfo{person}{Kai Dang}, \bibinfo{person}{Yang Fan}, \bibinfo{person}{Yichang Zhang}, \bibinfo{person}{An Yang}, \bibinfo{person}{Rui Men}, \bibinfo{person}{Fei Huang}, \bibinfo{person}{Bo Zheng}, \bibinfo{person}{Yibo Miao}, \bibinfo{person}{Shanghaoran Quan}, \bibinfo{person}{Yunlong Feng}, \bibinfo{person}{Xingzhang Ren}, \bibinfo{person}{Xuancheng Ren}, \bibinfo{person}{Jingren Zhou}, {and} \bibinfo{person}{Junyang Lin}.} \bibinfo{year}{2024}\natexlab{}.
\newblock \bibinfo{title}{Qwen2.5-Coder Technical Report}.
\newblock
\showeprint[arxiv]{2409.12186}~[cs.CL]
\urldef\tempurl%
\url{https://arxiv.org/abs/2409.12186}
\showURL{%
\tempurl}


\bibitem[Intel(2024a)]%
        {Intel-SPMD}
\bibfield{author}{\bibinfo{person}{Intel}.} \bibinfo{year}{2024}\natexlab{a}.
\newblock \bibinfo{title}{Intel® Implicit SPMD Program Compiler}.
\newblock \bibinfo{howpublished}{\url{https://ispc.github.io/}}.
\newblock


\bibitem[Intel(2024b)]%
        {Intel-Intrinsics-Guide}
\bibfield{author}{\bibinfo{person}{Intel}.} \bibinfo{year}{2024}\natexlab{b}.
\newblock \bibinfo{title}{Intel® Intrinsics Guide}.
\newblock \bibinfo{howpublished}{\url{https://www.intel.com/content/www/us/en/docs/intrinsics-guide/index.html}}.
\newblock


\bibitem[International(2025)]%
        {RVV-intrinsic}
\bibfield{author}{\bibinfo{person}{RISC-V International}.} \bibinfo{year}{2025}\natexlab{}.
\newblock \bibinfo{title}{RISC-V Vector Intrinsic Document}.
\newblock \bibinfo{howpublished}{\url{https://github.com/riscv-non-isa/rvv-intrinsic-doc}}.
\newblock


\bibitem[Jain et~al\mbox{.}(2025)]%
        {jain2025testgenevalrealworldunit}
\bibfield{author}{\bibinfo{person}{Kush Jain}, \bibinfo{person}{Gabriel Synnaeve}, {and} \bibinfo{person}{Baptiste Rozière}.} \bibinfo{year}{2025}\natexlab{}.
\newblock \bibinfo{title}{TestGenEval: A Real World Unit Test Generation and Test Completion Benchmark}.
\newblock
\showeprint[arxiv]{2410.00752}~[cs.SE]
\urldef\tempurl%
\url{https://arxiv.org/abs/2410.00752}
\showURL{%
\tempurl}


\bibitem[Jimenez et~al\mbox{.}(2024)]%
        {jimenez2024swebench}
\bibfield{author}{\bibinfo{person}{Carlos~E Jimenez}, \bibinfo{person}{John Yang}, \bibinfo{person}{Alexander Wettig}, \bibinfo{person}{Shunyu Yao}, \bibinfo{person}{Kexin Pei}, \bibinfo{person}{Ofir Press}, {and} \bibinfo{person}{Karthik~R Narasimhan}.} \bibinfo{year}{2024}\natexlab{}.
\newblock \showarticletitle{{SWE}-bench: Can Language Models Resolve Real-world Github Issues?}. In \bibinfo{booktitle}{\emph{Proceedings of the 12th International Conference on Learning Representations}}.
\newblock
\urldef\tempurl%
\url{https://openreview.net/forum?id=VTF8yNQM66}
\showURL{%
\tempurl}


\bibitem[Lai et~al\mbox{.}(2023)]%
        {ds1000}
\bibfield{author}{\bibinfo{person}{Yuhang Lai}, \bibinfo{person}{Chengxi Li}, \bibinfo{person}{Yiming Wang}, \bibinfo{person}{Tianyi Zhang}, \bibinfo{person}{Ruiqi Zhong}, \bibinfo{person}{Luke Zettlemoyer}, \bibinfo{person}{Wen-tau Yih}, \bibinfo{person}{Daniel Fried}, \bibinfo{person}{Sida Wang}, {and} \bibinfo{person}{Tao Yu}.} \bibinfo{year}{2023}\natexlab{}.
\newblock \showarticletitle{DS-1000: A Natural and Reliable Benchmark for Data Science Code Generation}. In \bibinfo{booktitle}{\emph{Proceedings of the 40th International Conference on Machine Learning}}.
\newblock
\urldef\tempurl%
\url{https://arxiv.org/abs/2211.11501}
\showURL{%
\tempurl}


\bibitem[Li et~al\mbox{.}(2025)]%
        {li2025tritonbenchbenchmarkinglargelanguage}
\bibfield{author}{\bibinfo{person}{Jianling Li}, \bibinfo{person}{Shangzhan Li}, \bibinfo{person}{Zhenye Gao}, \bibinfo{person}{Qi Shi}, \bibinfo{person}{Yuxuan Li}, \bibinfo{person}{Zefan Wang}, \bibinfo{person}{Jiacheng Huang}, \bibinfo{person}{Haojie Wang}, \bibinfo{person}{Jianrong Wang}, \bibinfo{person}{Xu Han}, \bibinfo{person}{Zhiyuan Liu}, {and} \bibinfo{person}{Maosong Sun}.} \bibinfo{year}{2025}\natexlab{}.
\newblock \bibinfo{title}{TritonBench: Benchmarking Large Language Model Capabilities for Generating Triton Operators}.
\newblock
\showeprint[arxiv]{2502.14752}~[cs.CL]
\urldef\tempurl%
\url{https://arxiv.org/abs/2502.14752}
\showURL{%
\tempurl}


\bibitem[Li et~al\mbox{.}(2022)]%
        {science2022codegen}
\bibfield{author}{\bibinfo{person}{Yujia Li}, \bibinfo{person}{David Choi}, \bibinfo{person}{Junyoung Chung}, \bibinfo{person}{Nate Kushman}, \bibinfo{person}{Julian Schrittwieser}, \bibinfo{person}{Rémi Leblond}, \bibinfo{person}{Tom Eccles}, \bibinfo{person}{James Keeling}, \bibinfo{person}{Felix Gimeno}, \bibinfo{person}{Agustin~Dal Lago}, \bibinfo{person}{Thomas Hubert}, \bibinfo{person}{Peter Choy}, \bibinfo{person}{Cyprien de Masson~d’Autume}, \bibinfo{person}{Igor Babuschkin}, \bibinfo{person}{Xinyun Chen}, \bibinfo{person}{Po-Sen Huang}, \bibinfo{person}{Johannes Welbl}, \bibinfo{person}{Sven Gowal}, \bibinfo{person}{Alexey Cherepanov}, \bibinfo{person}{James Molloy}, \bibinfo{person}{Daniel~J. Mankowitz}, \bibinfo{person}{Esme~Sutherland Robson}, \bibinfo{person}{Pushmeet Kohli}, \bibinfo{person}{Nando de Freitas}, \bibinfo{person}{Koray Kavukcuoglu}, {and} \bibinfo{person}{Oriol Vinyals}.} \bibinfo{year}{2022}\natexlab{}.
\newblock \showarticletitle{Competition-level Code Generation with AlphaCode}.
\newblock \bibinfo{journal}{\emph{Science}} (\bibinfo{year}{2022}).
\newblock
\href{https://doi.org/10.1126/science.abq1158}{doi:\nolinkurl{10.1126/science.abq1158}}


\bibitem[Liu et~al\mbox{.}(2023)]%
        {liu2023evalplus}
\bibfield{author}{\bibinfo{person}{Jiawei Liu}, \bibinfo{person}{Chunqiu~Steven Xia}, \bibinfo{person}{Yuyao Wang}, {and} \bibinfo{person}{Lingming Zhang}.} \bibinfo{year}{2023}\natexlab{}.
\newblock \showarticletitle{Is Your Code Generated by Chat{GPT} Really Correct? Rigorous Evaluation of Large Language Models for Code Generation}. In \bibinfo{booktitle}{\emph{Proceedings of the 37th International Conference on Neural Information Processing Systems}}.
\newblock
\urldef\tempurl%
\url{https://arxiv.org/abs/2305.01210}
\showURL{%
\tempurl}


\bibitem[Liu et~al\mbox{.}(2024a)]%
        {EvalPlus-Leaderboard}
\bibfield{author}{\bibinfo{person}{Jiawei Liu}, \bibinfo{person}{Songrun Xie}, \bibinfo{person}{Junhao Wang}, \bibinfo{person}{Yuxiang Wei}, \bibinfo{person}{Yifeng Ding}, {and} \bibinfo{person}{Lingming Zhang}.} \bibinfo{year}{2024}\natexlab{a}.
\newblock \bibinfo{title}{EvalPlus Leaderboard}.
\newblock \bibinfo{howpublished}{\url{https://evalplus.github.io/leaderboard}}.
\newblock


\bibitem[Liu et~al\mbox{.}(2024b)]%
        {evalperf2024}
\bibfield{author}{\bibinfo{person}{Jiawei Liu}, \bibinfo{person}{Songrun Xie}, \bibinfo{person}{Junhao Wang}, \bibinfo{person}{Yuxiang Wei}, \bibinfo{person}{Yifeng Ding}, {and} \bibinfo{person}{Lingming Zhang}.} \bibinfo{year}{2024}\natexlab{b}.
\newblock \showarticletitle{Evaluating Language Models for Efficient Code Generation}. In \bibinfo{booktitle}{\emph{Proceedings of the 1st Conference on Language Modeling}}.
\newblock
\urldef\tempurl%
\url{https://openreview.net/forum?id=IBCBMeAhmC}
\showURL{%
\tempurl}


\bibitem[Mendis et~al\mbox{.}(2019)]%
        {vectorization-nips19}
\bibfield{author}{\bibinfo{person}{Charith Mendis}, \bibinfo{person}{Cambridge Yang}, \bibinfo{person}{Yewen Pu}, \bibinfo{person}{Dr.Saman Amarasinghe}, {and} \bibinfo{person}{Michael Carbin}.} \bibinfo{year}{2019}\natexlab{}.
\newblock \showarticletitle{Compiler Auto-Vectorization with Imitation Learning}. In \bibinfo{booktitle}{\emph{Proceedings of the 33rd International Conference on Neural Information Processing Systems}}.
\newblock
\urldef\tempurl%
\url{https://dl.acm.org/doi/10.5555/3454287.3455597}
\showURL{%
\tempurl}


\bibitem[Microsoft(2025)]%
        {ONNXRuntime}
\bibfield{author}{\bibinfo{person}{Microsoft}.} \bibinfo{year}{2025}\natexlab{}.
\newblock \bibinfo{title}{ONNX Runtime}.
\newblock \bibinfo{howpublished}{\url{https://github.com/microsoft/onnxruntime}}.
\newblock


\bibitem[Nuzman and Henderson(2006)]%
        {vectorization-CGO06}
\bibfield{author}{\bibinfo{person}{Dorit Nuzman} {and} \bibinfo{person}{Richard Henderson}.} \bibinfo{year}{2006}\natexlab{}.
\newblock \showarticletitle{Multi-Platform Auto-Vectorization}. In \bibinfo{booktitle}{\emph{Proceedings of the 4th International Symposium on Code Generation and Optimization}}.
\newblock
\href{https://doi.org/10.1109/CGO.2006.25}{doi:\nolinkurl{10.1109/CGO.2006.25}}


\bibitem[Nuzman et~al\mbox{.}(2006)]%
        {vectorization-pldi06}
\bibfield{author}{\bibinfo{person}{Dorit Nuzman}, \bibinfo{person}{Ira Rosen}, {and} \bibinfo{person}{Ayal Zaks}.} \bibinfo{year}{2006}\natexlab{}.
\newblock \showarticletitle{Auto-Vectorization of Interleaved Data for SIMD}. In \bibinfo{booktitle}{\emph{Proceedings of the 27th ACM SIGPLAN Conference on Programming Language Design and Implementation}}.
\newblock
\href{https://doi.org/10.1145/1133981.1133997}{doi:\nolinkurl{10.1145/1133981.1133997}}


\bibitem[OpenAI(2024)]%
        {openai2024gpt4}
\bibfield{author}{\bibinfo{person}{OpenAI}.} \bibinfo{year}{2024}\natexlab{}.
\newblock \bibinfo{title}{GPT-4 Technical Report}.
\newblock
\showeprint[arxiv]{2303.08774}~[cs.CL]
\urldef\tempurl%
\url{https://arxiv.org/abs/2303.08774}
\showURL{%
\tempurl}


\bibitem[OpenAI(2025)]%
        {openai-platform}
\bibfield{author}{\bibinfo{person}{OpenAI}.} \bibinfo{year}{2025}\natexlab{}.
\newblock \bibinfo{title}{OpenAI Platform}.
\newblock \bibinfo{howpublished}{\url{https://platform.openai.com/docs/models}}.
\newblock


\bibitem[OpenCV.AI(2025)]%
        {OpenCV}
\bibfield{author}{\bibinfo{person}{OpenCV.AI}.} \bibinfo{year}{2025}\natexlab{}.
\newblock \bibinfo{title}{OpenCV}.
\newblock \bibinfo{howpublished}{\url{https://opencv.org/}}.
\newblock


\bibitem[Peng et~al\mbox{.}(2025)]%
        {peng2025coffe}
\bibfield{author}{\bibinfo{person}{Yun Peng}, \bibinfo{person}{Jun Wan}, \bibinfo{person}{Yichen Li}, {and} \bibinfo{person}{Xiaoxue Ren}.} \bibinfo{year}{2025}\natexlab{}.
\newblock \bibinfo{title}{COFFE: A Code Efficiency Benchmark for Code Generation}.
\newblock
\showeprint[arxiv]{2502.02827}~[cs.SE]
\urldef\tempurl%
\url{https://arxiv.org/abs/2502.02827}
\showURL{%
\tempurl}


\bibitem[Shypula et~al\mbox{.}(2024)]%
        {shypula2024pie}
\bibfield{author}{\bibinfo{person}{Alexander Shypula}, \bibinfo{person}{Aman Madaan}, \bibinfo{person}{Yimeng Zeng}, \bibinfo{person}{Uri Alon}, \bibinfo{person}{Jacob Gardner}, \bibinfo{person}{Milad Hashemi}, \bibinfo{person}{Graham Neubig}, \bibinfo{person}{Parthasarathy Ranganathan}, \bibinfo{person}{Osbert Bastani}, {and} \bibinfo{person}{Amir Yazdanbakhsh}.} \bibinfo{year}{2024}\natexlab{}.
\newblock \bibinfo{title}{Learning Performance-Improving Code Edits}.
\newblock
\showeprint[arxiv]{2302.07867}~[cs.SE]
\urldef\tempurl%
\url{https://arxiv.org/abs/2302.07867}
\showURL{%
\tempurl}


\bibitem[simdjson(2025)]%
        {simdjson}
\bibfield{author}{\bibinfo{person}{simdjson}.} \bibinfo{year}{2025}\natexlab{}.
\newblock \bibinfo{title}{simdjson}.
\newblock \bibinfo{howpublished}{\url{https://github.com/simdjson/simdjson}}.
\newblock


\bibitem[Siso et~al\mbox{.}(2019)]%
        {vectorization-eval-taco19}
\bibfield{author}{\bibinfo{person}{Sergi Siso}, \bibinfo{person}{Wes Armour}, {and} \bibinfo{person}{Jeyarajan Thiyagalingam}.} \bibinfo{year}{2019}\natexlab{}.
\newblock \showarticletitle{Evaluating Auto-Vectorizing Compilers through Objective Withdrawal of Useful Information}.
\newblock \bibinfo{journal}{\emph{ACM Transactions on Architecture and Code Optimization}} (\bibinfo{year}{2019}).
\newblock
\href{https://doi.org/10.1145/3356842}{doi:\nolinkurl{10.1145/3356842}}


\bibitem[Team(2025a)]%
        {yang2025qwen251mtechnicalreport}
\bibfield{author}{\bibinfo{person}{Qwen Team}.} \bibinfo{year}{2025}\natexlab{a}.
\newblock \bibinfo{title}{Qwen2.5-1M Technical Report}.
\newblock
\showeprint[arxiv]{2501.15383}~[cs.CL]
\urldef\tempurl%
\url{https://arxiv.org/abs/2501.15383}
\showURL{%
\tempurl}


\bibitem[Team(2025b)]%
        {qwen2025qwen25technicalreport}
\bibfield{author}{\bibinfo{person}{Qwen Team}.} \bibinfo{year}{2025}\natexlab{b}.
\newblock \bibinfo{title}{Qwen2.5 Technical Report}.
\newblock
\showeprint[arxiv]{2412.15115}~[cs.CL]
\urldef\tempurl%
\url{https://arxiv.org/abs/2412.15115}
\showURL{%
\tempurl}


\bibitem[Theodoridis and Su(2024)]%
        {RefinedInputDegradedOutput}
\bibfield{author}{\bibinfo{person}{Theodoros Theodoridis} {and} \bibinfo{person}{Zhendong Su}.} \bibinfo{year}{2024}\natexlab{}.
\newblock \showarticletitle{Refined Input, Degraded Output: The Counterintuitive World of Compiler Behavior}. In \bibinfo{booktitle}{\emph{Proceedings of the 45th ACM SIGPLAN Conference on Programming Language Design and Implementation}}.
\newblock
\href{https://doi.org/10.1145/3656404}{doi:\nolinkurl{10.1145/3656404}}


\bibitem[xAI(2025)]%
        {grok}
\bibfield{author}{\bibinfo{person}{xAI}.} \bibinfo{year}{2025}\natexlab{}.
\newblock \bibinfo{title}{Grok}.
\newblock \bibinfo{howpublished}{\url{https://grok.com/}}.
\newblock


\bibitem[Yang et~al\mbox{.}(2025)]%
        {yang2024swebenchmultimodal}
\bibfield{author}{\bibinfo{person}{John Yang}, \bibinfo{person}{Carlos~E. Jimenez}, \bibinfo{person}{Alex~L. Zhang}, \bibinfo{person}{Kilian Lieret}, \bibinfo{person}{Joyce Yang}, \bibinfo{person}{Xindi Wu}, \bibinfo{person}{Ori Press}, \bibinfo{person}{Niklas Muennighoff}, \bibinfo{person}{Gabriel Synnaeve}, \bibinfo{person}{Karthik~R. Narasimhan}, \bibinfo{person}{Diyi Yang}, \bibinfo{person}{Sida~I. Wang}, {and} \bibinfo{person}{Ofir Press}.} \bibinfo{year}{2025}\natexlab{}.
\newblock \showarticletitle{{SWE}-bench Multimodal: Do AI Systems Generalize to Visual Software Domains?}. In \bibinfo{booktitle}{\emph{Proceedings of the 13th International Conference on Learning Representations}}.
\newblock
\urldef\tempurl%
\url{https://openreview.net/forum?id=riTiq3i21b}
\showURL{%
\tempurl}


\bibitem[Yu et~al\mbox{.}(2024)]%
        {codereval2024}
\bibfield{author}{\bibinfo{person}{Hao Yu}, \bibinfo{person}{Bo Shen}, \bibinfo{person}{Dezhi Ran}, \bibinfo{person}{Jiaxin Zhang}, \bibinfo{person}{Qi Zhang}, \bibinfo{person}{Yuchi Ma}, \bibinfo{person}{Guangtai Liang}, \bibinfo{person}{Ying Li}, \bibinfo{person}{Qianxiang Wang}, {and} \bibinfo{person}{Tao Xie}.} \bibinfo{year}{2024}\natexlab{}.
\newblock \showarticletitle{CoderEval: A Benchmark of Pragmatic Code Generation with Generative Pre-trained Models}. In \bibinfo{booktitle}{\emph{Proceedings of the IEEE/ACM 46th International Conference on Software Engineering}}.
\newblock
\href{https://doi.org/10.1145/3597503.3623316}{doi:\nolinkurl{10.1145/3597503.3623316}}


\bibitem[Zheng et~al\mbox{.}(2024)]%
        {multilingual2024codegeex}
\bibfield{author}{\bibinfo{person}{Qinkai Zheng}, \bibinfo{person}{Xiao Xia}, \bibinfo{person}{Xu Zou}, \bibinfo{person}{Yuxiao Dong}, \bibinfo{person}{Shan Wang}, \bibinfo{person}{Yufei Xue}, \bibinfo{person}{Zihan Wang}, \bibinfo{person}{Lei Shen}, \bibinfo{person}{Andi Wang}, \bibinfo{person}{Yang Li}, \bibinfo{person}{Teng Su}, \bibinfo{person}{Zhilin Yang}, {and} \bibinfo{person}{Jie Tang}.} \bibinfo{year}{2024}\natexlab{}.
\newblock \bibinfo{title}{CodeGeeX: A Pre-Trained Model for Code Generation with Multilingual Benchmarking on HumanEval-X}.
\newblock
\showeprint[arxiv]{2303.17568}~[cs.LG]
\urldef\tempurl%
\url{https://arxiv.org/abs/2303.17568}
\showURL{%
\tempurl}


\bibitem[Zheng et~al\mbox{.}(2025)]%
        {zheng2025vectrans}
\bibfield{author}{\bibinfo{person}{Zhongchun Zheng}, \bibinfo{person}{Kan Wu}, \bibinfo{person}{Long Cheng}, \bibinfo{person}{Lu Li}, \bibinfo{person}{Rodrigo C.~O. Rocha}, \bibinfo{person}{Tianyi Liu}, \bibinfo{person}{Wei Wei}, \bibinfo{person}{Jianjiang Zeng}, \bibinfo{person}{Xianwei Zhang}, {and} \bibinfo{person}{Yaoqing Gao}.} \bibinfo{year}{2025}\natexlab{}.
\newblock \bibinfo{title}{VecTrans: Enhancing Compiler Auto-Vectorization through LLM-Assisted Code Transformations}.
\newblock
\showeprint[arxiv]{2503.19449}~[cs.SE]
\urldef\tempurl%
\url{https://arxiv.org/abs/2503.19449}
\showURL{%
\tempurl}


\bibitem[Zhuo et~al\mbox{.}(2025)]%
        {zhuo2025bigcodebench}
\bibfield{author}{\bibinfo{person}{Terry~Yue Zhuo}, \bibinfo{person}{Minh~Chien Vu}, \bibinfo{person}{Jenny Chim}, \bibinfo{person}{Han Hu}, \bibinfo{person}{Wenhao Yu}, \bibinfo{person}{Ratnadira Widyasari}, \bibinfo{person}{Imam Nur~Bani Yusuf}, \bibinfo{person}{Haolan Zhan}, \bibinfo{person}{Junda He}, \bibinfo{person}{Indraneil Paul}, \bibinfo{person}{Simon Brunner}, \bibinfo{person}{Chen Gong}, \bibinfo{person}{Thong Hoang}, \bibinfo{person}{Armel~Randy Zebaze}, \bibinfo{person}{Xiaoheng Hong}, \bibinfo{person}{Wen-Ding Li}, \bibinfo{person}{Jean Kaddour}, \bibinfo{person}{Ming Xu}, \bibinfo{person}{Zhihan Zhang}, \bibinfo{person}{Prateek Yadav}, \bibinfo{person}{Naman Jain}, \bibinfo{person}{Alex Gu}, \bibinfo{person}{Zhoujun Cheng}, \bibinfo{person}{Jiawei Liu}, \bibinfo{person}{Qian Liu}, \bibinfo{person}{Zijian Wang}, \bibinfo{person}{Binyuan Hui}, \bibinfo{person}{Niklas Muennighoff}, \bibinfo{person}{David Lo}, \bibinfo{person}{Daniel Fried}, \bibinfo{person}{Xiaoning Du}, \bibinfo{person}{Harm de
  Vries}, {and} \bibinfo{person}{Leandro~Von Werra}.} \bibinfo{year}{2025}\natexlab{}.
\newblock \bibinfo{title}{BigCodeBench: Benchmarking Code Generation with Diverse Function Calls and Complex Instructions}.
\newblock
\showeprint[arxiv]{2406.15877}~[cs.SE]
\urldef\tempurl%
\url{https://arxiv.org/abs/2406.15877}
\showURL{%
\tempurl}


\end{thebibliography}

\end{document}